\documentclass[12pt,draftcls,journal,a4paper,twoside,onecolumn]{IEEEtran}

\ifCLASSOPTIONcompsoc
  \usepackage[nocompress]{cite}
\else
  \usepackage{cite}
\fi

\ifCLASSINFOpdf
  \usepackage[pdftex]{graphicx}
\else
  \usepackage[dvips]{graphicx}
\fi

\usepackage[cmex10]{amsmath}
\usepackage{algorithmic}
\usepackage{array}
\usepackage{mdwmath}
\usepackage{mdwtab}
\usepackage{eqparbox}

\ifCLASSOPTIONcompsoc
  \usepackage[caption=false,font=normalsize,labelfont=sf,textfont=sf]{subfig}
\else
  \usepackage[caption=false,font=footnotesize]{subfig}
\fi

\usepackage{stfloats}
\usepackage{fixltx2e}
\usepackage{url}

\usepackage{amsbsy}
\usepackage{amssymb}
\usepackage{amscd}
\usepackage{amsmath}

\newtheorem{theorem}{Theorem}
\newtheorem{corollary}{Corollary}

\def\K{{\mathcal K}}
\def\A{{\mathcal A}}
\def\C{{\mathcal C}}
\def\T{{\mathcal T}}

\begin{document}

\date{March 17, 2012}

\title{On Optimal Link Activation with Interference Cancellation in Wireless Networking}

\author{Di~Yuan,~\IEEEmembership{Member,~IEEE,}
        Vangelis~Angelakis,~\IEEEmembership{Member,~IEEE,}
        Lei~Chen,~\IEEEmembership{Student~Member,~IEEE,}
        Eleftherios~Karipidis,~\IEEEmembership{Member,~IEEE,}
        and~Erik~G.~Larsson,~\IEEEmembership{Senior~Member,~IEEE}
\thanks{D. Yuan, V. Angelakis, and L. Chen are with the Mobile Telecommunications Group, Department of Science and Technology (ITN), Link{\"o}ping University, SE-601 74 Norrk{\"o}ping, Sweden
(e-mail: diyua@itn.liu.se, vangelis.angelakis@liu.se, leich@itn.liu.se).}
\thanks{E. Karipidis and E. G. Larsson are with the Communication Systems Division, Department of Electrical Engineering (ISY), Link\"oping University, SE-581 83 Link\"oping, Sweden
(e-mail:  karipidis@isy.liu.se, erik.larsson@isy.liu.se).}
}

\markboth{S\MakeLowercase{ubmitted to} \textit{IEEE Transactions on Vehicular
Technology} \MakeLowercase{on }M\MakeLowercase{arch 17, 2012}}{}

\maketitle

\begin{abstract}
A fundamental aspect in performance engineering of wireless networks is 
optimizing the set of links that can be concurrently activated to meet given 
signal-to-interference-and-noise ratio (SINR) thresholds. 
The solution of this combinatorial problem
is the key element in scheduling and cross-layer resource
management. Previous works on link activation assume
single-user decoding receivers, that treat interference
in the same way as noise. In this paper, we assume
multiuser decoding receivers, which can cancel
strongly interfering signals. As a result, in
contrast to classical spatial reuse, links being close to each other
are more likely to be active simultaneously. Our goal here is to deliver a comprehensive
theoretical and numerical study on optimal link activation under this novel setup, 
in order to provide insight into the gains from adopting interference cancellation.
We therefore consider the optimal problem setting of successive interference cancellation (SIC), 
as well as the simpler, yet instructive, case of parallel interference cancellation (PIC).  
We prove that both problems are NP-hard and develop compact integer linear programming
formulations that enable us to approach the global optimum solutions.
We provide an extensive numerical performance evaluation,
indicating that for low to medium SINR thresholds  
the improvement  is quite substantial, especially with SIC, whereas for high SINR thresholds
the improvement diminishes and both schemes perform equally well.
\end{abstract}


%
\IEEEpeerreviewmaketitle

\section{Introduction}\label{sec:introduction}\par

In wireless networking, determining the sets of links that can be active
simultaneously is a cornerstone optimization task of combinatorial
nature. For a link to be active, a given
signal-to-interference-and-noise ratio (SINR) threshold must be met at
the receiver, according to the physical connectivity model
\cite{GupK00}. Within this domain, previous analyses assume that the
communication system employs single-user decoding (SUD) receivers
that treat interference as additive noise. For interference-limited scenarios, it is very
unlikely that all links can be active at the same time. Hence, it is
necessary to construct transmission schedules that orthogonalize link
transmissions along some dimension of freedom, such as time. The
schedule is composed by link subsets, each of which is a feasible
solution to the link activation (LA) problem. Thus, for scheduling,
repeatedly solving the LA problem becomes the dominant
computational task. Intuitively, with SUD, a solution to the LA
problem consists in links being spatially separated, as they generate
little interference to each other. Thus, scheduling amounts to optimal
spatial reuse of the time resource. For this reason, scheduling is
also referred to as spatial time-division multiple access (STDMA) \cite{NeKl85}.
Optimal LA has attracted a considerable amount of attention. Problem complexity and
solution approximations have been addressed in
\cite{GoPsWa07,AnDi09,BraBS06,XuT09}. A recent algorithmic
advance is presented in \cite{CaChGuYu11}. Research on scheduling,
which uses LA as the building block, is extensive; see,
e.g, \cite{BjVaYu04,CaCaFiGuMa10,Pant09} and references
therein. In addition to scheduling, LA is an integral part of more complicated resource
management problems jointly addressing scheduling and other resource
control aspects, such as rate adaptation and power control, as well as
routing, in ad hoc and mesh networks; see, e.g.,
\cite{CaCaFiGuMa10,LiEp07,ElAsJa10}.

In the general problem setting of LA, each link is
associated with a nonnegative weight, and the objective is to
maximize the total weight of the active links. The weights may be used
to reflect utility values of the links or queue sizes
\cite{Tass92}. A different view of weights comes from the column generation,
method proposed in \cite{BjVaYu03,BjVaYu04}, which has become the standard solution
algorithm for scheduling as well as for joint scheduling,
power control, and rate adaptation \cite{CaCa06}.
The algorithm decomposes the problem to a master problem
and a subproblem, both of which are much more tractable than the
original. Solving the subproblem constructs a feasible LA
set. In the subproblem, the links are associated with prices coming
from the linear programming dual, corresponding to the weights of our
LA problem. A special case of the weights is a vector of
ones; in this case, the objective becomes to maximize the cardinality
of the LA set.

All aforementioned previous works on optimal LA have assumed SUD,
for which interference is regarded as additive noise.
In this work, we examine the problem of optimal LA under a novel setup;
namely when the receivers have multiuser decoding (MUD) capability
\cite{Verdu1998}. Note that, unlike noise, interference contains
encoded information and hence is a structured signal. This is exploited
by MUD receivers to perform interference cancellation (IC).
That is, the receivers, before decoding the signal of interest,
first decode the interfering signals they are able to and
remove them from the received signal. For IC to take place, a receiver acts as
though it is the intended receiver of the interfering signal. Therefore,
an interfering signal can be cancelled, i.e., decoded at the rate it
was actually transmitted, only if it is received with enough power in
relation to the other transmissions, including the receiver's
original signal of interest.
In other words, the ``interference-to-other-signals-and-noise'' ratio
(which is an intuitive but non-rigorous term in this context), must meet the SINR threshold of the
interfering signal. With MUD, the effective SINR of the signal of interest is
higher than the original SINR, with SUD, since the
denominator now only contains the sum of the residual,
i.e., undecoded, interference plus noise. Clearly, with MUD,
concurrent activation of strongly interfering
links becomes more likely, enabling activation patterns that are
counter-intuitive in the conventional STDMA setting. The focus of our
investigation is on the potential of IC in boosting the performance of
LA. Because LA is a key element in many
resource management problems, the investigation opens up new
perspectives of these problems as well.

The topic of implementing MUD receivers in real systems has recently gained
interest, particularly in the low SINR domain using low-complexity
algorithms; see, e.g., \cite{Ghaffar2009}. Technically, implementing IC
is not a trivial task. A fundamental assumption in MUD is that the
receivers have information (codebooks and modulation schemes) of the
transmissions to be cancelled. Furthermore, the transmitters need to
be synchronized in time and frequency. Finally, the
receivers must estimate, with sufficient accuracy, the channels
between themselves and all transmitters whose signals are trying to
decode. For our work, we assume that MUD is carried out without any
significant performance impairments, and examine it as an enabler of
going beyond the conventionally known performance limits in wireless networking. Hence,
the results we provide effectively constitute upper bounds on what can
be achievable, for the considered setup, in practice.

The significance of introducing MUD and more specifically IC to
wireless networking is motivated by the fundamental, i.e., information-theoretic,
studies of the so-called interference channel, which accurately models the physical-layer
interactions of the transmissions on coupled links. The capacity region of the
interference channel is a long-standing open problem, even for the
two-link case, dating back to \cite{Ahlswede1974,Carleial1978,Han1981}.
Up to now, it is only known in a few special cases; see, e.g.,
\cite{Shang2009,Annapureddy2009} for some recent contributions. Two
basic findings, regarding optimal treatment of interference in the
two-link case, can be summarized as follows.  When the interference is
very weak, it can simply be treated as additive noise.  When the
interference is strong enough, it may be decoded and subtracted off
from the received signal, leaving an interference-free signal
containing only the signal of interest plus thermal noise.

From a physical-layer perspective, the simple
two-link setting above corresponds to a received signal consisting of
$X=S+I+N$, where $S$ is the signal of interest, with received power $P_S$ and
encoded with rate $R_S$, $I$ is the interference signal with received power
$P_I$ encoded with rate $R_I$, and $N$ is the receiver noise with power
$\eta$. Assuming Gaussian signaling and capacity achieving codes, the interference $I$ is ``strong enough'' to be decoded,
treating the signal of interest $S$ as additive noise, precisely if
\begin{equation}\label{eq:intro:example1}
 \log_2\left(1 + \frac{P_I}{P_S + \eta}\right) \ge R_I \Leftrightarrow \frac{P_I}{P_S + \eta} \geq \gamma_I,
\end{equation}
where $\gamma_I \triangleq 2^{R_I} -1$ denotes the SINR threshold for decoding the interference signal $I$.
If condition (\ref{eq:intro:example1}) holds, i.e., the ``interference-to-other-signal-and-noise'' ratio is at least $\gamma_I$,
$I$ can be decoded perfectly and subtracted off from $X$. Then, decoding the signal of interest $S$ is possible, provided
that the interference-free part of $X$ has sufficient signal-to-noise ratio (SNR)
\begin{equation}\label{eq:intro:example2}
  \log_2\left(1 + \frac{P_S}{\eta}\right) \ge R_S \Leftrightarrow \frac{P_S}{\eta} \geq \gamma_S,
\end{equation}
where $\gamma_S \triangleq 2^{R_S} -1$ denotes the SINR threshold for decoding signal $S$.
By contrast, if the interference is not sufficiently strong for
(\ref{eq:intro:example1}) to hold, then it must be treated as additive noise.
In such a case, decoding of signal $S$ is possible only when
\begin{equation}\label{eq:intro:example3}
 \frac{P_S}{P_I + \eta} \geq \gamma_S.
\end{equation}

This way of reasoning can be extended to more than one interfering
signals. Towards this end, we examine the effect of IC
in scenarios with potentially many links in
transmission. Our study has a clear focus on performance engineering
in wireless networking with arbitrary topologies. In consequence, a
thorough study of the gain of IC to LA is highly
motivated, in view of the pervasiveness of the LA problem
in resource management of many types of wireless networks.
In the multi-link setup that we consider, the optimal scheme is to
allow every receiver perform IC successively, i.e., in multiple stages.
In every stage, the receiver decodes one interfering signal,
removes it from the received signal,
and continues as long as there exists an undecoded interfering link
whose received signal is strong enough in relation to the sum of the
residual interference, the signal of interest, and noise. This scheme
is referred to as \emph{successive IC (SIC)}.
From an optimization standpoint, modeling SIC mathematically is very
challenging, because the order in which cancellations take place is of
significance. Clearly, enumerating the potential cancellation orders
will not scale at all. Thus compact formulations that are able to
deliver the optimal order are essential, especially under the physical connectivity model,
which quantifies interference accurately.

Alternatively, a simplified IC scheme is to consider only the cancellations that can be performed concurrently, in a single stage.
In this scheme, when determining the possibility for the cancellation of an
interfering link, all remaining transmissions, no matter whether or
not they are also being examined for cancellation, are
regarded as interference. We refer to this scheme as \emph{parallel IC
(PIC)}. It is easily realized that some of the cancellations in SIC may
not be possible in PIC; thus one can expect that the gain of the
latter is less than that of the former. A further restriction is to
allow at most one cancellation per receiver. This scheme, which we
refer to as \emph{single-link IC (SLIC)}, poses additional limit on the
performance gain. However, it is the simplest scheme for
practical implementation and frequently captures most of the performance gain due to IC.
In comparison to SIC, PIC and SLIC are much
easier to formulate mathematically, as ordering is not relevant.

In \cite{Karipidis2011}, we evaluated the potential of SLIC in the
related problem of SINR balancing.  That is, we considered as input
the number of active links, let the transmit powers be variables and
looked for the maximum SINR level that can be guaranteed to all links.
In \cite{Angelakis11b}, we exploited link rate adaptation to maximize the
benefits of IC to aggregate system throughput.  In parallel, another
set of authors has made a relevant contribution in the context of IC
\cite{Lv11}.  They considered a SIC-enabled system and introduced a
greedy algorithm to construct schedules of bounded length in ad-hoc
networks with MUD capabilities.  There though, the interference is
modeled using the protocol-based model of conflict graphs \cite{Ja03},
which simplifies the impact of interference, in comparison to the more
accurate physical connectivity model that we are using.  


The overall aim of our work is to deliver a comprehensive
theoretical and numerical study on optimal link activation under this novel setup in order to provide insight into the gains from adopting interference cancellation.
This is achieved through the following contributions: 
\begin{itemize}
\item First, we
introduce and formalize the optimization problems of LA in wireless
networks with PIC and SIC, focusing on the latter most challenging
case. 
\item Second, we prove that these optimization problems
are NP-hard. 
\item Third, we develop ILP formulations that enable us to
approach the global optimum for problem sizes of practical interest
and thus provide an effective benchmark for the potential of IC on
LA.
\item Fourth, we present an extensive numerical
performance evaluation that introduces insight into the maximum attainable gains
of adopting IC. 
\end{itemize}
We show that for some of the test scenarios the improvement is
substantial. The results indicate that strong interference can indeed
be taken as a great advantage in designing new notions for scheduling
and cross-layer resource allocation in wireless networking with MUD
capabilities.

The remainder of the paper is organized as follows. In Section~\ref{sec:preliminaries},
we introduce the notation and formalize the
novel optimization problems. In Section~\ref{sec:complexity},
 we prove the theoretical complexity results.
In Section~\ref{sec:singlestage}, we propose a compact ILP
formulation for the LA problem with PIC having quadratic size to the number of links.
For the most challenging problem of LA with SIC, we devote two sections. In
Section~\ref{sec:sicuniform}, we treat SIC under a common SINR
threshold. By exploiting the problem structure, we show that the order of cancellations
can be conveniently modeled and derive an
ILP formulation of quadratic size. 
Then, in Section \ref{sec:sicvariable},
we consider individual SINR thresolds. For this case, we give an
ILP formulation of cubic size.
In Section \ref{sec:simulation}, we present and discuss simulation
results evaluating the performance of all proposed IC schemes.
In Section \ref{sec:conclusion}, we give conclusions and outline
perspectives.

\section{Definition of Link Activation with Interference Cancellation}
\label{sec:preliminaries}

Consider a wireless system of $K$ pairs of transmitters and receivers,
forming $K$ directed links.  The discussions in the forthcoming
sections can be easily generalized to a network where the nodes can act as
both transmitters or receivers.  Let $\K \triangleq \{1, \dots, K\}$
denote the set of links.
The gain of the channel between the transmitter of link $m$ and the
receiver of link $k$ is denoted by $G_{mk}$, for any two $m,k \in
\K$.  The noise power is denoted by $\eta$ and, for simplicity,
is assumed equal at all receivers. The SINR threshold of link $k$ is
denoted by $\gamma_k$. Each link $k$ is associated with a predefined
positive activation weight $w_k$, reflecting its utility value or queue size
or dual price. If a link is activated, its transmit
power is given and denoted by $p_k$, for $k \in \K$.
A LA set is said to be feasible if the SINR thresholds of the
links in the set are met under simultaneous transmission.
All versions of the LA problems we consider have the same
input that we formalize below.\\
\textbf{Input:} A link set $\K$ with the following parameters:
transmit powers $p_{k}$, SINR thresholds $\gamma_{k}$, and link
weights $w_{k}$, $\forall k \in \K$, and gain values $G_{mk}$,
$\forall m, k \in \K$.

Consider first the LA problem with the conventional assumption of SUD,
where the interference is treated as additive noise. This is the
baseline version of the LA problem in our comparisons; its output
is formulated as follows.

\underline{\textbf{Problem LA-SUD:}} \emph{Optimal link activation with single-user decoding.}\\
\textbf{Output:} An activation set $\A \subseteq \K$, maximizing $\sum_{k \in \A}\limits w_k$ and satisfying the conditions:\\[1ex]
\begin{equation}
\frac{p_k G_{kk}} {\sum_{m\in\A \setminus \{k\}}\limits p_m G_{mk} + \eta } \geq \gamma_k \qquad \forall k \in \A.
\end{equation}

This classical version of the LA problem can be
represented by means of an ILP formulation; see, e.g.,
\cite{BjVaYu04,CaChGuYu11}. A set of binary variables $x_{k}, \forall k \in \K$,
is used to indicate whether or not each of the links is active. The
activation set is hence $\A = \{k \in \K: ~x_k = 1\}$. In order to ease comparisons to the
formulations that are introduced later, we
reproduce below the formulation of LA-SUD:
\begin{subequations}\label{eq:prelim:model}\begin{align}
\label{eq:prelim:objective}
\max~~ & \sum_{k \in \K} w_kx_k \\[1ex]
\text{s. t.}~~ &  \frac{p_k G_{kk} + M_{k}(1 - x_{k})}{\sum_{m \not=k}\limits p_m G_{mk} x_m + \eta} \geq \gamma_k  \qquad \forall k \in \K, \label{eq:prelim:SINR}\\
& x_k \in \{0, 1\} \qquad \forall k \in \K. \label{eq:prelim:x_k01}
\end{align}\end{subequations}

The objective function \eqref{eq:prelim:objective} aims to maximize
the total weight of the LA set.  The constraints
\eqref{eq:prelim:SINR} formulate the SINR
criteria. If $x_{k}=1$, indicating that link $k$ is active, the $k$th
inequality constrains the SINR of link $k$ to be at least $\gamma_k$.
For the case that link $k$ is not active, $x_{k}=0$, the $k$th
inequality in
\eqref{eq:prelim:SINR} is always satisfied, i.e., it has null effect, if
parameter $M_k$ is set to a sufficiently large value. By
construction, an obvious choice is $M_{k} =
\sum_{m\not=k}\limits p_m G_{mk}\gamma_k +
\eta\gamma_k - p_k G_{kk}$. Note that the size of the formulation \eqref{eq:prelim:model},
both in the numbers of variables and constraints, is of $O(K)$.

Now, consider the same system but with receivers having MUD capability that enable
cancellation of strongly interfering links. We distinguish between
IC in a single stage (PIC) and in multiple stages (SIC).
In the former, to cancel the transmission of an interfering
link, all other signals of active links, including the signal of interest, are
considered to be additive noise, independent of other cancellation
decisions at the same receiver. A formal definition of the output
is given below.

\underline{\textbf{Problem LA-PIC:}} \emph{Optimal link activation with parallel interference cancellation.}\\
\textbf{Output:} An activation set $\A \subseteq \K$ and
the set $\C_k \subseteq {\cal{A}} \setminus \{k\}$ of cancelled transmissions for each $k \in \A$, maximizing
$\sum_{k \in \A}\limits w_k$ and satisfying the conditions:
\begin{subequations}\begin{align}
&\frac{p_m G_{mk}} {\sum_{n \in \A \setminus {\{m\}} }\limits p_n G_{nk} + \eta } \geq \gamma_m \qquad \forall m \in \C_k, ~\forall k \in \A, \label{eq:pic}\\[1ex]
&\frac{p_k G_{kk}}{\sum_{m\in\A \setminus\{k,\C_k\} }\limits p_m G_{mk} + \eta } \geq \gamma_k \qquad \forall k \in \A. \label{eq:sinr1}
\end{align}\end{subequations}

The set of conditions \eqref{eq:pic} ensures that the specified cancellations can take
place. For the receiver of link $k$ to cancel the transmission of
link $m$, the receiver of $k$ acts as if it was the receiver of $m$.
Hence, the ``interference-to-other-signals-and-noise'' ratio incorporates in the numerator
the received power $p_mG_{mk}$ of the interfering link $m$  and in the denominator
the received power $p_kG_{kk}$ of own link $k$.
This ratio must satisfy the SINR threshold of the signal $m$ to be decoded.
The set of conditions \eqref{eq:sinr1} formulates the SINR
requirements for the signals of interest, taking into account the effect
of IC in the SINR ratio. That is,  the cancelled terms are removed from the sum in the denominator,
determining the aggregate power of the undecoded interference which is treated as additive noise.

For SIC, the output must be augmented in order to specify,
in addition to the cancellations $\C_k$, by the receiver of $k \in \A$, the
order in which they take place. A formal definition of the output is
given below.

\underline{\textbf{Problem LA-SIC:}} \emph{Optimal link activation with successive interference cancellation.}\\
\textbf{Output:} An activation set $\A \subseteq \K$ and
the set $\C_k \subseteq \A \setminus \{k\}$ of cancelled transmissions
along with a bijection \mbox{$b_k: \C_k \mapsto \{1,\dots, |\C_k| \}$} for each $k\in \A$, maximizing $\sum_{k \in \A}\limits w_k$ and satisfying the conditions:
\begin{subequations}
\begin{equation}
\frac{p_m G_{mk}}{\sum_{n \in \A \setminus
\{m,q\in\C_k:~b_k(q)<b_k(m)\}}\limits p_n G_{nk} + \eta} \geq \gamma_m
\qquad \forall m \in \C_k,~\forall k \in \A, \label{eq:sic}
\end{equation}
\begin{equation}
\frac{p_k G_{kk}}{\sum_{m\in\A \setminus\{k,\C_k\}}\limits p_m G_{mk} + \eta } \geq \gamma_k
\qquad \forall k \in \A.  \qquad\qquad \label{eq:sinr2}
\end{equation}
\end{subequations}

Similarly to LA-PIC, the set of conditions \eqref{eq:sic} formulates
the requirement for SIC and the set \eqref{eq:sinr2} the requirement
for decoding the signals of interest, taking into account the effect
of IC in the SINR ratio. In the output, the cancellation sequence for
each $k$ in the activation set is given by the bijection $b_k$; the
bijection defines a unique mapping of the link indices in the
cancellation set $\C_k$ to the IC order numbers in the cancellation
sequence.  That is, $b_k(m), m \in \C_k$ defines the stage at which
link $m$ is cancelled by the receiver of link $k$.  The bijection is
used in the IC conditions \eqref{eq:sic}, in order to exclude from the
sum in the denominator, the interference terms that have been
cancelled in stages prior to $m$.

\section{Complexity}
\label{sec:complexity}

The baseline problem, LA-SUD, is known to be NP-hard; see, e.g., \cite{GoPsWa07}.
For a combinatorial optimization problem,
introducing new elements to the problem structure may change the
complexity level, potentially making the problem easier to
solve. Hence, without additional investigation, the NP-hardness of
LA-SUD does not carry over to LA with IC. In this
section, we provide the theoretical result that problems LA-PIC and LA-SIC remain NP-hard,
using a unified proof applicable to both cases.

\begin{theorem}
\label{theorem:hardness}
Problem LA-PIC is NP-hard.
\end{theorem}

\begin{IEEEproof}
We provide a reduction from LA-SUD to LA-PIC. Considering an
arbitrary instance of LA-SUD, we construct an instance of LA-PIC as follows.
For each link $k \in \K$, we go through all other
links in $\K
\setminus \{k\}$ one by one. Let $m$ be the link under
consideration. The power of link $k$ is set to
\begin{equation}
\label{eq:proofpower}
p'_k \triangleq \max \left\{p_k, \left(\displaystyle\frac{p_m G_{mk}}{\gamma_m - \varepsilon}-\eta \right) / G_{kk} \right\},
\end{equation}
where $\varepsilon$ is a small positive constant. By
\eqref{eq:proofpower}, the power of $k$ is either kept as before, or grows by an
amount such that
$\displaystyle\frac{p_m G_{mk}} {p'_k G_{kk} + \eta } < \gamma_m$.
Therefore, link $k$ is not able to
decode the signal of $m$, i.e., the IC condition of LA-PIC cannot
be satisfied, even in the most favorable scenario that all other links, apart from $m$ and $k$, are inactive.

After any power increase of link $k$, we make sure that this update
does not have any effect in the application of
\eqref{eq:proofpower} to the other links. This is achieved by scaling down
the channel gain $G_{km}$ as
\begin{equation} \label{eq:proofgain}
G'_{km} \triangleq G_{km}p_k /p'_k,  
\end{equation}
meaning that for any $m$, the received signal strength from $k$
remains the same as in the original instance of LA-SUD.
As a result, even though IC is allowed in the instance of LA-PIC, no
cancellation will actually take place, since none of the IC conditions
holds due to the scalings in \eqref{eq:proofpower} and \eqref{eq:proofgain}.

By the construction above, for each link $k \in \K$ the total
interference that is treated as noise in the instance of LA-PIC equals
that in the instance of LA-SUD. Thus, the denominator of the SINR of the signal of interest
does not change. On the other hand, the numerator may have grown from
$p_k$ to $p'_k$. To account for this growth, the
SINR threshold $\gamma_k$ is set to
\begin{equation} \label{eq:proofgamma}
\gamma'_k \triangleq \gamma_k p_k /p'_k.  
\end{equation}
In effect, the increase of the power on a link, if any, is compensated for by the new SINR threshold.
Note that, because $p'_k / p_k \geq 1$, $\gamma'_k$ prohibits cancellation of the $k$th signal by any receiver other than the $k$th one.

From the construction, one can conclude that a LA set is
feasible in the instances of LA-SUD, if and only if this is the
case in the instance of LA-PIC. In addition, the reduction is
clearly polynomial.  Hence the conclusion.
\end{IEEEproof}

\begin{corollary}
\label{th:SIChardness}
Problem LA-SIC is NP-hard.
\end{corollary}

\begin{IEEEproof}
The result follows immediately from the fact that, in the proof of
Theorem \ref{theorem:hardness}, the construction does not impose any
restriction on the number of links to be cancelled, nor to the order
in which the cancellations take place.
\end{IEEEproof}

\section{Link Activation with Parallel Interference Cancellation}
\label{sec:singlestage}

In this section, we propose a compact ILP formulation for LA-PIC.
In addition to the $x_k,~\forall k \in \K$, variables in \eqref{eq:prelim:model}, we
introduce a second set of binary variables, $y_{mk},~\forall m, k \in \K, ~m
\not=k$. Variable $y_{mk}$ is one if the receiver of link $k$ decodes
and cancels the interference from link $m$ and zero otherwise. The output
of LA-PIC is then defined by $\A = \{ k \in \K: x_k =
1\}$ and $\C_k = \{m \in \A\setminus\{k\}:~y_{mk} = 1\}$, for each $k \in \A$. The proposed formulation for LA-PIC is

\begin{subequations}
\label{eq:SSIC_model}
\begin{align}
\max~~~ &\sum_{k \in \K} {w_kx_k}\label{eq:SSIC_objective}\\
\text{s. t.~~} & y_{mk} \leq x_m \qquad \forall m, k \in \K, ~ m \neq k, \label{eq:SSIC_activ_cond_m} \\[1ex]
& y_{mk} \leq x_k \qquad \forall m, k \in \K, ~ m \neq k, \label{eq:SSIC_activ_cond_k} \\[1ex]
& \frac{p_k G_{kk} + M_{k}(1 - x_{k})}{\sum_{m \neq k}\limits p_m G_{mk} (x_m -y_{mk}) + \eta} \geq \gamma _k \qquad \forall k \in \K, \label{eq:SSIC_SINR_k} \\[1ex]
& \frac{p_m G_{mk} + M_{mk}(1 - y_{mk})}{\sum_{n \neq m}\limits p_n G_{nk} x_n + \eta} \geq \gamma_m
\qquad \forall m, k \in \K, ~ m \neq k, \label{eq:SSIC_ISNR_m} \\[1ex]
& y_{mk} \in \{0, 1\} \qquad \forall m, k \in \K, ~ m \neq k,\\[1ex]
& x_{k} \in \{0, 1\} \qquad \forall k \in \K.
\end{align}\end{subequations}

The objective function \eqref{eq:SSIC_objective} is the same as
\eqref{eq:prelim:objective} for LA-SUD. The first two sets of inequalities,
\eqref{eq:SSIC_activ_cond_m} and \eqref{eq:SSIC_activ_cond_k}, pose necessary
conditions on the relation between the variable values. Namely, a cancellation
can take place, i.e., $y_{mk}=1$, only if both links $k$ and $m$ are active,
i.e., $x_k=x_m=1$. The set of inequalities \eqref{eq:SSIC_SINR_k} formulates
the SINR requirements \eqref{eq:sinr1} for decoding the signals of interest, in
a way similar to \eqref{eq:prelim:SINR} for LA-SUD, with the difference that
here the cancelled interference terms are subtracted from the denominator using
the term $x_m - y_{mk}$. Note that, without \eqref{eq:SSIC_activ_cond_m}, the
formulation will fail, as in \eqref{eq:SSIC_SINR_k} it would allow to reduce
the denominator of the ratio by subtracting non-existing interference from
non-active links. The next set of constraints \eqref{eq:SSIC_ISNR_m} formulates
the condition \eqref{eq:pic} for PIC: $y_{mk}$ can be set to one only if the
interference from link $m$, $p_m G_{mk}$, is strong enough in relation to all
other active signals, including the signal of interest. If the ratio meets the
SINR threshold $\gamma_m$ for link $m$, cancellation can be carried out.
Setting $y_{mk}$ to zero is always feasible, on the other hand, provided that
the parameter $M_{mk}$ is large enough. A sufficiently large value is $M_{mk}
\triangleq \sum_{n \neq m}\limits p_n G_{nk}\gamma_m + \eta\gamma_m - p_m
G_{mk}$. The construction of \eqref{eq:SSIC_ISNR_m} reflects the fact of
performing all cancellations in a single stage, as in cancelling the signal of
link $m$, other transmissions being cancelled in parallel are treated as
additive noise. Note that the model remains in fact valid even if
\eqref{eq:SSIC_activ_cond_k} is removed. Doing so would allow the receiver of
an inactive link to perform IC. However, since an inactive link does not
contribute at all to the objective function, this is a minor ``semantic''
mismatch that can be simply alleviated by post-processing.

For practical purposes, each receiver may be allowed to cancel the signal of at most one interfering link.
The resulting LA problem, denoted LA-SLIC, can be easily formulated by adapting the formulation \eqref{eq:SSIC_model} for LA-PIC.
The only required change is the addition of the set of constraints
\begin{equation}\label{eq:SSIC_SLIC}
\sum_{m \neq k} y_{mk} \leq 1 \qquad \forall k \in \K,
\end{equation}
that restricts each receiver to cancel at most one interfering transmission.

The size of the formulation \eqref{eq:SSIC_model}, both in the numbers of variables and constraints, is
 of $O(K^2)$. Thus, the formulation is compact and its size grows by one magnitude in
comparison to \eqref{eq:prelim:model}. In fact, to incorporate
cancellation between link pairs, one cannot expect any optimization
formulation of smaller size.

When implementing the formulation, two pre-processing steps can be applied to
reduce the size of the problem and hence speed-up the calculation of the solution.
First, the links that are infeasible, taking into account only the receiver noise, are identified
by checking for every receiver whether the received SNR meets the SINR threshold for activation.
If the answer is ``false'', i.e., $\displaystyle\frac{p_k G_{kk}} {\eta } < \gamma_k$,
then link $k$ is removed from consideration by fixing the $x_k$ variable to zero.
Second, the link pairs for which cancellation can never take place are found by checking,
for every receiver and interfering signal, whether the ``interference-to-signal-of-interest-and-noise'' ratio
meets the SINR threshold for decoding the interference signal.
If the answer is ``false'', i.e., $\displaystyle\frac{p_m G_{mk}} {p_k G_{kk} + \eta } < \gamma_m$,
then link $k$ cannot decode the interference from $m$ and this option is eliminated from the formulation by
setting the respective variable $y_{mk}$ to zero.

\section{Link Activation with Successive Interference Cancellation Under Common SINR Threshold}
\label{sec:sicuniform}

Incorporating the optimal IC scheme, SIC, to the LA problem is highly desired,
since it may activate sets that are infeasible by PIC. However, using
ILP to formulate compactly the solution space of LA-SIC, is challenging.
This is because the formulation has to deal, for each link, with a bijection giving the cancellation
sequence. We propose an ILP approach and present it in
two steps. In this section, we consider LA-SIC under the assumption that all links
have a common SINR threshold for activation, i.e., $\gamma_k=\gamma, ~\forall k \in
\K$. In the next section, we address the general case of individual SINR thresholds.

For SIC under common SINR threshold, we exploit the problem structure and show
that the optimal cancellation order can be handled implicitly in the
optimization formulation. As a result, we show that LA-SIC can in fact be
formulated as compactly as LA-PIC, i.e., using $O(K^2)$ variables and
constraints. The idea is to formulate an optimality condition on the ordering
of IC. To this end, consider an arbitrary link $k$ and observe that meeting the
SINR threshold for decoding the signal of interest is equivalent to having in
the receiver, after IC, a total amount of undecoded interference and noise at
most equal to $p_k G_{kk} / \gamma$. We refer to this term as the
\emph{interference margin} $u_{k}$ . Similarly, the interference margin that
allows cancellation of the interference from link $m$ at the receiver of link
$k$ is $u_{mk} \triangleq p_m G_{mk} / \gamma$. Consider any two interfering
links $m$ and $n$, and suppose $u_{mk} > u_{nk}$. Note that, because the SINR
threshold is common, the condition is equivalent to $p_m G_{mk} > p_n G_{nk}$,
i.e., the receiver of link $k$ experiences stronger interference from $m$. If
the condition holds, the cancellation of $m$ should be ``easier'' in some
sense. Thus, one may expect that if $k$ can decode both $m$ and $n$, the
decoding of $m$ should take place first. In the following, we prove a theorem,
stating that this is indeed the case at the optimum---there exists an optimal
solution having the structure in which if a weaker interference signal can be
cancelled, then any stronger one is cancelled before it.

\begin{theorem}
\label{th:order}
If $u_{mk} > u_{nk}$ and the receiver of link $k$ is able to cancel the signal of
$n$, then it is feasible to cancel the signal of $m$ before
$n$ and there exists at least one optimum having this structure
in the cancellation sequence.
\end{theorem}

\begin{IEEEproof}
Let $I_{nk}$ denote the total power of undecoded interference and noise
when the receiver of $k$ decodes the signal from $n$.
Assume that $m$ has not been cancelled in a previous stage.
Then, $p_m G_{mk}$ is part of $I_{nk}$.
Successful cancellation of $n$ means that $I_{nk} \leq u_{nk}$.
Since $u_{nk} < u_{mk}$, it holds that $I_{nk} < u_{mk}$.
Consider now decoding the signal of $m$ immediately before $n$.
Thus for this cancellation, the total power of the undecoded
interference and noise incorporates the interference of $n$, but not that of $m$,
i.e., $I_{mk} = I_{nk} + p_n G_{nk} - p_m G_{mk}$.
Because $u_{mk} > u_{nk}$ implies $p_m G_{mk} > p_n G_{nk}$, it holds that
$I_{mk} < I_{nk}$. Since $I_{nk} < u_{mk}$, the cancellation
condition $I_{mk} \leq u_{mk}$ is satisfied. After cancelling $m$,
IC can still take place for $n$, because the new $I_{nk}$ is decreased by $p_m G_{mk}$.
Consequently, both $m$ and $n$ can be cancelled. Obviously,
doing so will not reduce the number of active links and
the theorem follows.
\end{IEEEproof}

By Theorem~\ref{th:order}, for each link $k$, we can perform a pre-ordering
of all other links in descending order of their interference margins. SIC at link $k$ can be restricted to this order without loss
of optimality. At the optimum, the cancellations performed by $k$, for
interfering links that are active, will follow the order, until no
more additional cancellations can take place. In this optimal
solution, when considering the cancellation condition of interfering link $m$,
interference can only originate from links appearing after $m$ in the
sorted sequence.

We propose an ILP formulation based on Theorem~\ref{th:order}.  The
formulation uses the same variables of \eqref{eq:SSIC_model} for
LA-PIC, as there is no need to formulate the cancellation order
explicitly. The sorted sequence is denoted by, for each link $k\in\K$,
a bijection $i_k: \K \setminus
\{k\} \mapsto \{1, \dots, K-1\}$, where $i_k(m)$ is the position of link $m$
in the sorted sequence. The sorting results in $i_k(m) > i_k(n)$ if
$u_{mk} < u_{nk}$. In case of $u_{mk} = u_{nk}$, the tie can be broken
arbitrarily without affecting the optimization result. In addition,
let $c_{mk} \triangleq K -1 - i_k(m)$ denote the number of links
appearing after $m$ in the sorted sequence for $k$. The proposed
formulation for LA-SIC under common SINR threshold is

\begin{subequations}
\label{eq:ModelIc}
\begin{align}
\max~~ & \sum_{k \in \K} w_{k}x_{k}  \label{eq:ObjIc}\\
\text{s. t.}~~ & y_{mk} \leq x_m \qquad \forall m, k \in \K, ~ m \neq k,  \label{eq:Cond_mk} \\[1ex]
& y_{mk} \leq x_k \qquad \forall m, k \in \K, ~ m \neq k,  \label{eq:Cond_k} \\[1ex]
& \frac{p_k G_{kk} + M_{k}(1 - x_{k})}{\sum_{m \neq k}\limits p_m G_{mk} (x_m -y_{mk}) + \eta} \geq \gamma \qquad ~ \forall k \in \K, \label{eq:SINR_k} \\[1ex]
& \frac{p_m G_{mk} + M_{mk}(1 - y_{mk})}{\sum _{n \neq k, ~ i_{k}(n) >
i{_{k}(m)}} \limits p_n G_{nk} x_n + p_kG_{kk} + \eta} \geq \gamma
\qquad \forall m, k \in \K, ~ m \neq k,  \label{eq:SINR_mk} \\[1ex]
& \sum _{n \neq k, ~ i_{k}(n) > i{_{k}(m)}} y_{nk} \leq c_{mk} (1 - x_m +
y_{mk}) \qquad \forall m, k \in \K, ~ m \neq k,  \label{eq:ordering} \\[1ex]
& y_{mk} \in \{0, 1\} \qquad \forall m, k \in \K, ~ m \neq k, \\[1ex]
& x_{k} \in \{0, 1\} \qquad \forall k \in \K .
\end{align}\end{subequations}

The first three constraint sets \eqref{eq:Cond_mk}--\eqref{eq:SINR_k}
have the same meaning with \eqref{eq:SSIC_activ_cond_m}--\eqref{eq:SSIC_SINR_k}
 for LA-PIC; see Section~\ref{sec:singlestage}.
The constraint set (\ref{eq:SINR_mk}) formulates the conditions  \eqref{eq:sic} for SIC,
making use of Theorem~\ref{th:order}.
Consider the condition for cancellation of signal $m$ from receiver $k$ in stage $i_k(m)$.
Then, in the denominator of the ratio,
the sum of undecoded interference is limited to the transmissions coming after $m$ in the
sorted sequence of $k$, since all other active links with higher interference margin
than $m$ have already been cancelled.
The formulation is however not
complete without \eqref{eq:ordering}. This set of constraints, in
fact, ensures the optimality condition set by Theorem~\ref{th:order} and utilized
in \eqref{eq:SINR_mk}. That is, if both $m$ and $n$ are active, $u_{mk} >
u_{nk}$, and $n$ is cancelled by $k$, then $m$ is cancelled by $k$ as
well. Equivalently speaking, if $m$ is active but not cancelled by
$k$, then none of the other links after $m$ in the sequence of $k$ may
be cancelled. Examining \eqref{eq:ordering}, we see that it has no
effect as long as $x_m$ equals $y_{mk}$. If link $m$ is active but not
cancelled, corresponding to $x_m = 1$ and $y_{mk}=0$, the right-hand
side of \eqref{eq:ordering} becomes zero, and therefore no
cancellation will occur for any $n$ having position after $m$ in the ordered sequence.
Also, note that the case $x_m = 0$ but $y_{mk}=1$ cannot occur, because of
\eqref{eq:Cond_mk}.

Given a solution to the formulation \eqref{eq:ModelIc}, the cancellation
sequence of each active link $k$, i.e., the bijection $b_k$ in the definition of LA-SIC in Section~\ref{sec:preliminaries}, is
easily obtained by retrieving from the predefined bijection $i_k$
the elements with $y_{km}=1$.
The compactness of the formulation \eqref{eq:ModelIc} is manifested by the fact that its size, in both the numbers of
variables and constraints, is of $O(K^2)$. Thus, provided that there is a common SINR
threshold for activation, we have formulated LA-SIC as compactly as LA-PIC.

When implementing the formulation \eqref{eq:ModelIc}, similar pre-processing steps with \eqref{eq:SSIC_model} for LA-PIC can be applied to reduce the size of the problem. First, the infeasible links are removed for consideration by fixing $x_k$ to zero when $u_k < \eta$. Second, the infeasible IC options are eliminated from the formulation by fixing $y_{mk}$ to zero when $u_{mk} < p_kG_{kk} + \eta$.

\section{Link Activation with Successive Interference Cancellation Under Individual SINR Thresholds}
\label{sec:sicvariable}

In this section, we consider the LA-SIC problem under
the most general setup; namely when the links have individual SINR
thresholds. Differently from the common SINR case, treated in Section~\ref{sec:sicuniform},
a pre-ordering of the sequence of potential IC does not apply. The reason is that the
interference margin $u_{mk}$ does not depend anymore only on the received
power $p_mG_{mk}$ but also on the link-specific SINR threshold $\gamma_m$.
To see this point, consider a scenario where link $k$ attempts to cancel
the signal of two
interfering links $m$ and $n$ in two consecutive stages.  Denote by
$I$ the sum of the remaining interference, other than $m$ or $n$, the
received power of link $k$'s own signal, and noise. Assume a mismatch
between the relation of interference margin and that of received
power: $p_{m}G_{mk} < p_n G_{nk}$ but $u_{mk} > u_{nk}$ because
$\gamma_m < \gamma_n$. If $k$ cancels $m$ and then $n$, the
cancellation conditions are $u_{mk}
\geq p_n G_{nk} + I$ and $u_{nk} \geq I$. Reversing the cancellation order leads to
the conditions $u_{nk} \geq p_m G_{mk} + I$ and $u_{mk} \geq I$.  For our
example, we let $I = 0.5$. Consider two sets of values for the other
parameters. The values in the first set are: $u_{mk} = 3$, $u_{nk} = 1$, $p_m G_{mk}=1$, $p_nG_{nk}=2$
and in the second set are: $u_{mk} = 2$, $u_{nk} = 1$, $p_m G_{mk}=0.5$
$p_nG_{nk}=2$.
For the first set, both interfering links can be cancelled only
if cancellation applies to $m$ first, whereas the opposite order
must be used for the second set. Hence the interference margin (or received
power) does not provide a pre-ordering for cancellation.

In the following, we propose an ILP formulation for LA-SIC under individual SINR thresholds,
that explicitly accounts for the cancellation order.
Our approach is to introduce for each pair of links, $m,k\in\K$, a set of binary variables
$y_{mk}^t$ and represent the cancellation stage by the superscript $t$.
Variable $y_{mk}^t$ is
one if and only if the receiver of link $k$ cancels the interference from link $m$
at stage $t$. The effect is that, for each link $k$, the solution values of $y_{mk}^t$ order
the feasible cancellations; hence, they have a direct correspondence to the
output bijection $b_k$ of LA-SIC, defined in Section~\ref{sec:preliminaries}.
It is apparent that the index $t$ ranges between one and $K-1$.
In practice, due to computational considerations, we may want to restrict
the maximum number of cancellation stages. To this end, we define, for each $k\in\K$,
the integer parameters $T_k \leq K-1$ and the sets $\T_k \triangleq \{1,\dots,T_k\}$.
The proposed formulation of the general
LA-SIC problem, under individual SINR thresholds and restricted cancellation stages, is

\begin{subequations}
\label{eq:FinalModel}
\begin{align}
& \max~~ \sum_{k \in \K} w_{k}x_{k} \label{eq:FinalMax}\\[1ex]
& \text{s. t.} \nonumber \\
& \sum _{t=1}^{T_k} y^{t}_{mk}\leq x_m, \qquad \forall m, k \in \K, ~ m \neq k, \label{eq:FinalActiv}\\[1ex]
& \sum_{m \neq k} y^{t}_{mk} \leq x_k, \qquad \forall k \in \K, ~\forall t\in\T_k, \label{eq:Finalstage}\\[1ex]
& \frac{p_k G_{kk} + M_{k}(1 - x_{k})}{\sum_{m \not=k}\limits p_m G_{mk} \big(x_m -{\sum _{t=1}^{T_k}\limits y^{t}_{mk}}\big) + \eta} \geq \gamma_k  \qquad \forall k \in \K, \label{eq:FinalSINR} \\[1ex]
& \frac{p_m G_{mk} + M_{mk} (1-y_{mk}^t)}{\sum_{n \not=m,k }\limits p_n G_{nk} \big(x_n - { \sum_{t'=1}^{t-1} \limits y^{t'}_{nk}} \big) + p_k G_{kk} + \eta} \geq \gamma_m
\qquad \forall m, k \in \K, ~ m \neq k, ~\forall t\in\T_k, \label{eq:FinalISNR} \\[1ex]
& \sum_{m \neq k} \limits y^{t}_{mk} \leq \sum_{m \neq k}\limits y^{t-1}_{mk} \qquad \forall k \in \K, ~\forall t \in \T_k \setminus\{1\}, \label{eq:Finalblock}\\[1ex]
& y^{t}_{mk} \in \{0, 1\} \qquad \forall m, k \in \K, ~ m \neq k, ~\forall t\in\T_k, \label{eq:FINALy}\\[1ex]
& x_{k} \in \{0, 1\} \qquad \forall k \in \K.\label{eq:FINALx}
\end{align}\end{subequations}

The conditions \eqref{eq:FinalActiv}--\eqref{eq:Finalstage} have similar role with \eqref{eq:Cond_mk}--\eqref{eq:Cond_k}.
Namely, only when links $k$ and $m$ are active, the receiver of $k$ can consider to cancel the transmission of $m$.
In addition, the summation over $t$ in \eqref{eq:FinalActiv} ensures that link $m$ is cancelled in at most one stage.
Furthermore, the summation over $m$ in \eqref{eq:Finalstage} enforces each receiver to perform at most one cancellation per stage.
This removes equivalent solutions, without compromizing optimality, to enhance computational efficiency.
The SINR requirements for decoding the signals of interest are  set in \eqref{eq:FinalSINR}, similarly to \eqref{eq:SINR_k}.
In the denominator, all cancelled links, regardless of the stage the
cancellation is performed, are removed from the sum of undecoded interference.
The next set of constraints \eqref{eq:FinalISNR} formulates
the requirement for the cancellation of link $m$ by link $k$ at
stage $t$.
The active interfering transmissions that have been cancelled before stage $t$
are excluded from the sum of undecoded interference in the denominator of the ratio.
The constraints \eqref{eq:FinalISNR} are formulated with the convention that
the sum within the parenthesis in the denominator of the ratio is zero for $t=1$.
Note that, even though for each receiver $k$ and interfering link $m$, $T_k$ constraints are formulated,
due to \eqref{eq:FinalActiv}, all but at most one will be trivially satisfied by the respective $y_{mk}^t$ variabls being equal to zero.
The constraints \eqref{eq:Finalblock} are not mandatory for the correctness of the
formulation, but their role is to enhance the computational
efficiency. These constraints ensure that the cancellations
are performed as ``early'' as possible, i.e., there are no ``idle'' stages at which no cancellation takes place and which are followed by later stages where cancellation takes place. Otherwise, if $m_1$ and $m_2$ are
cancelled by $k$, and the cancellation of the former takes place first,
the cancellations can be performed at any two stages $t_1$ and $t_2$, as long as
$t_1<t_2$. Clearly, such solutions are all equivalent to each other,
and the presence of them slows down the computational process.

\begin{figure*}[tbp]
  \centering
  \subfloat[I dataset; sparse topology]{\label{fig:setup20spIT}\includegraphics[width=0.4\textwidth]{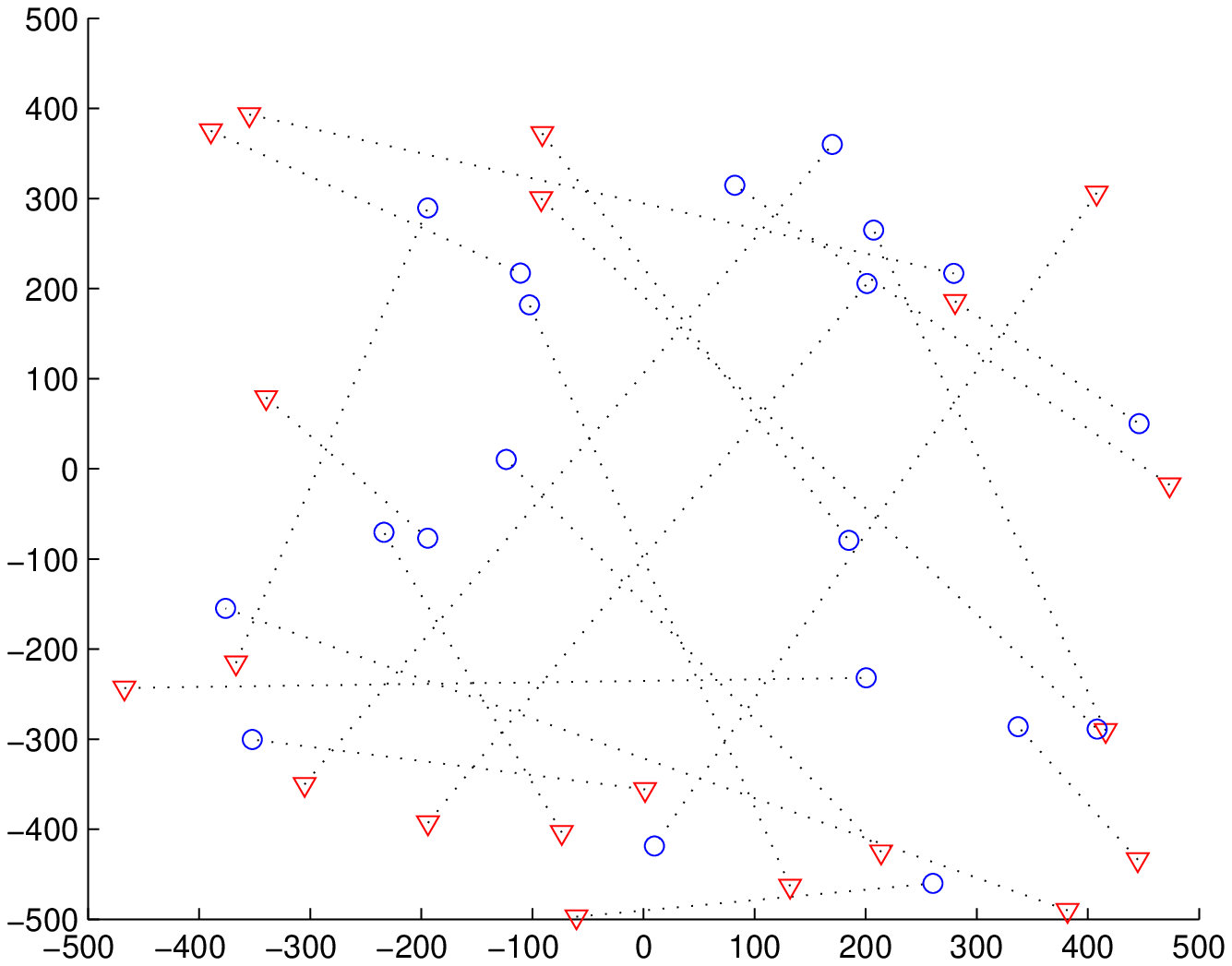}}
  \subfloat[N dataset; sparse topology]{\label{fig:setup20spNW}\includegraphics[width=0.4\textwidth]{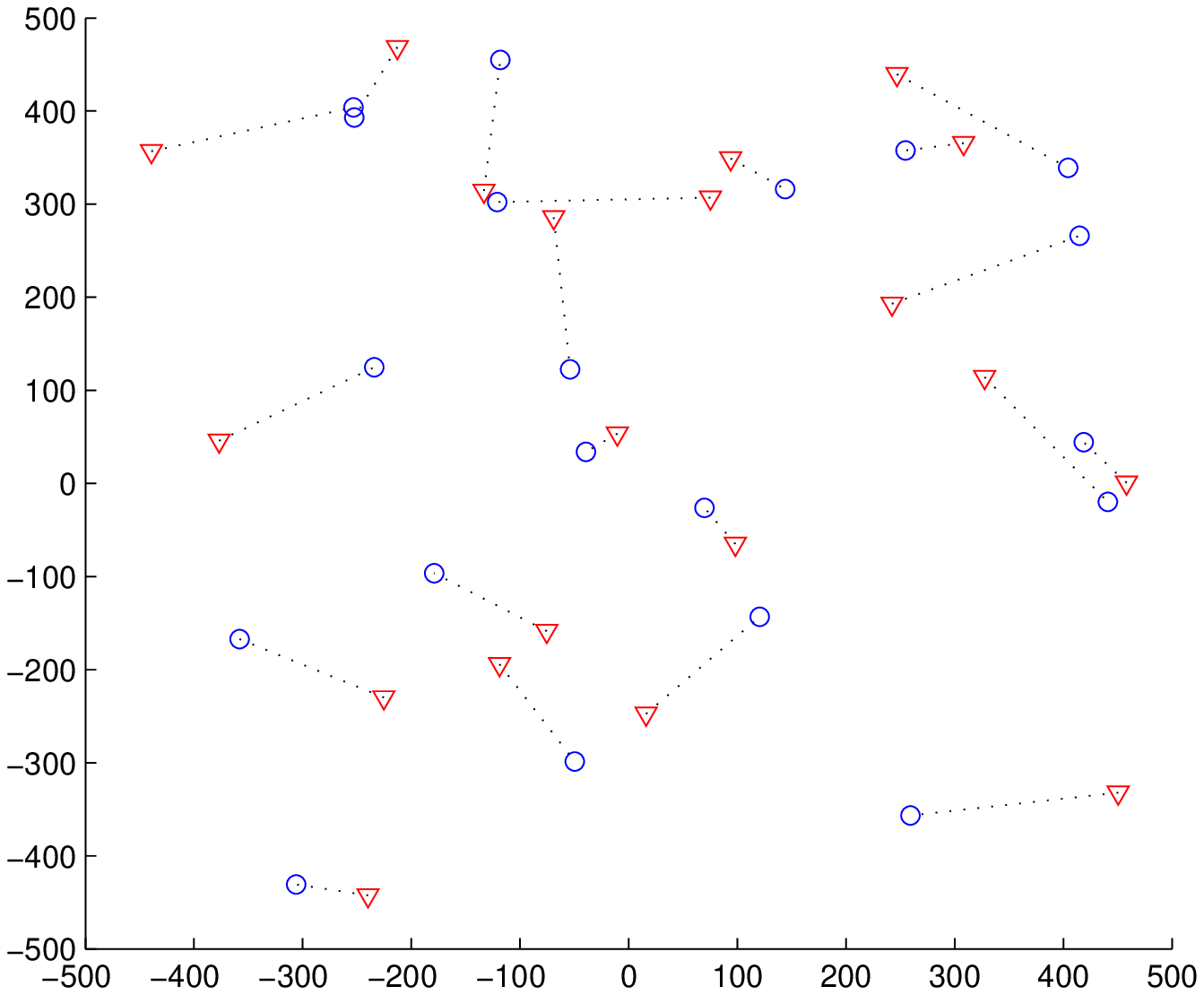}}\\
  \subfloat[I dataset; dense topology]{\label{fig:setup20deIT}\includegraphics[width=0.4\textwidth]{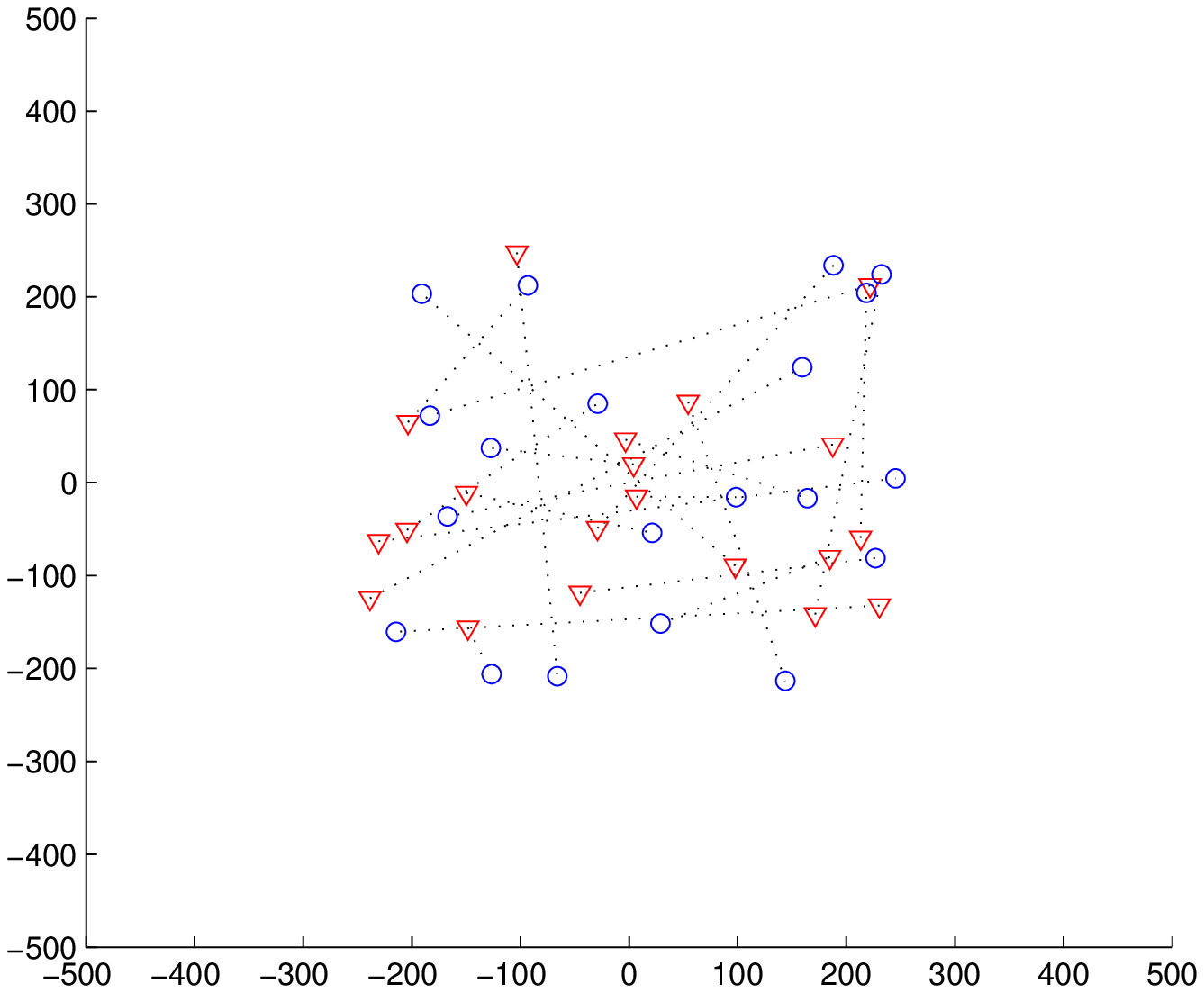}}
  \subfloat[N dataset; dense topology]{\label{fig:setup20deNW}\includegraphics[width=0.4\textwidth]{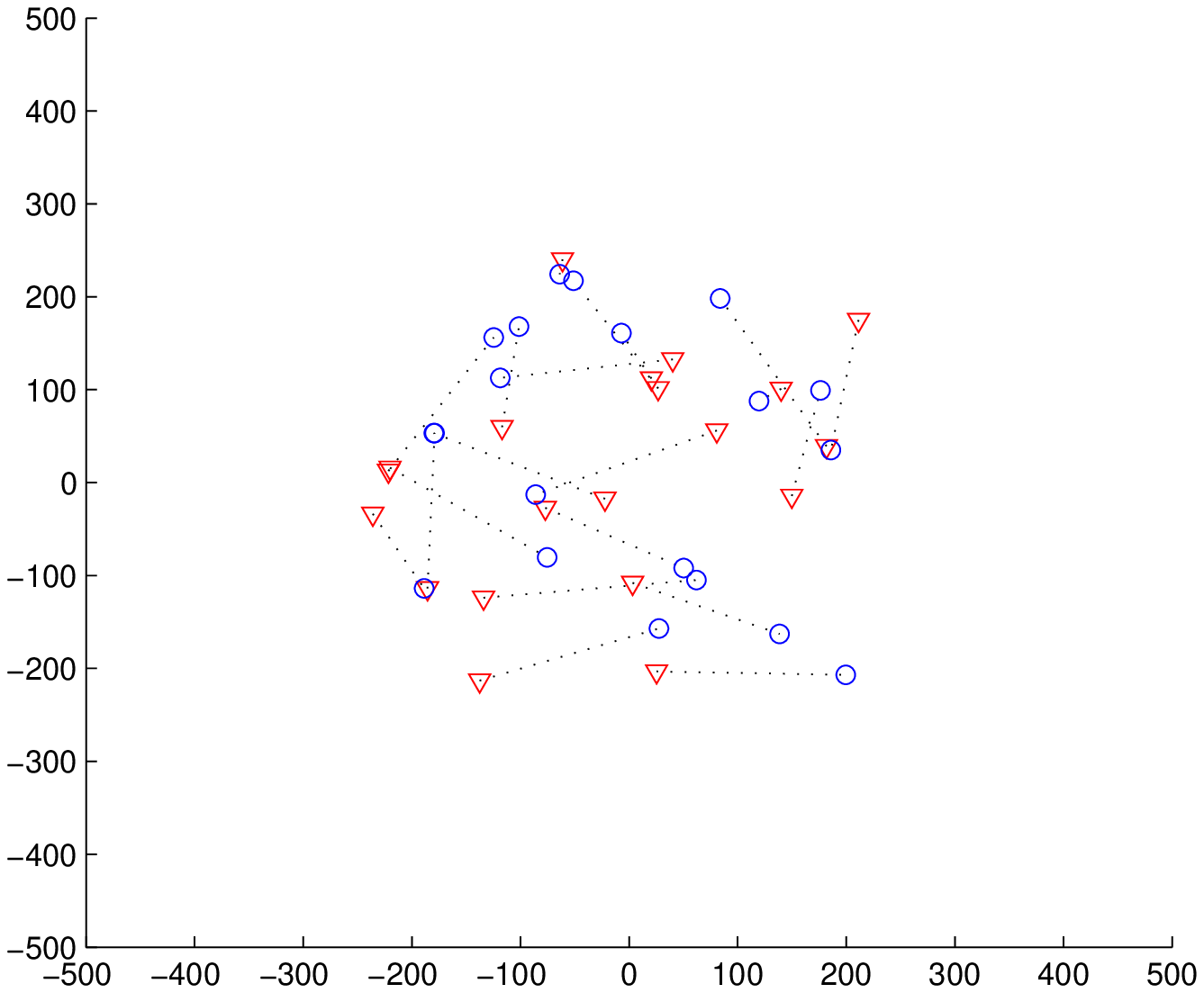}}
  \caption{Instances of a $20$-link network for different datasets and densities; transmitters marked with circles, receivers with triangles.}
  \label{fig:instances}
\end{figure*}

Since \eqref{eq:FinalModel} formulates the most general LA-SIC problem, it
also applies to the common-SINR case of Section~\ref{sec:sicuniform}.
Its computational efficiency though is significantly lower than the respective of formulation \eqref{eq:ModelIc}.
The reason is that its size is one magnitude larger than \eqref{eq:ModelIc},
i.e., the numbers of variables and constraints grow from $O(K^2)$ to $O(K^3)$.
However, we note that the formulation \eqref{eq:FinalModel} remains compact. In order to deal with
the scalability issue, one may resort to restrict the maximum
number of cancellations, $T_k$, to a constant being considerably lower
than $K-1$. Typically, doing so has little impact on the solution
quality, because most of the performance gain from IC is due to the
first few cancellations. Also, when implementing the formulation \eqref{eq:FinalModel},
similar pre-processing steps with \eqref{eq:ModelIc} can be applied, see Section~\ref{sec:sicuniform}, to reduce the size of the problem.

\section{Numerical results}
\label{sec:simulation}

This section presents a quantitative study of the effect that IC has
on the optimal LA problem in wireless networking. The ILP
formulations, proposed in Sections
\ref{sec:singlestage}--\ref{sec:sicvariable}, are utilized to conduct
extensive simulation experiments on randomly generated network
instances with various topologies, densities, cardinalities, and SINR
thresholds. Nodes are uniformly scattered in square areas of
$1000\text{m}\times 1000$m and $500\text{m} \times 500$m, in order to
create sparse and dense topologies, respectively. Two types of
datasets are generated. The first one takes an information-theoretic
viewpoint and is henceforth denoted dataset I. To this end, the
transmitter-receiver matchings are arbitrarily chosen \cite{BorEph06,
Hae05}, with the sole criterion of feasible single-link activation.
Thus, the links have arbitrary length within the test area, provided
that their SNR is larger than the SINR threshold required for
activation. The second dataset provides a rather networking-oriented
approach
\cite{FuLiHu2010, SriHae10} and is henceforth denoted dataset N. In
this dataset, the length of the links is constrained to be from $3$m up to
$200$m, with the rationale to produce instances resembling a multihop
network. The networks considered have cardinality $K$ ranging from $5$
up to $30$ links. Fig.~\ref{fig:instances} illustrates instances of a
20-link network; Figs.~\ref{fig:setup20spIT} and \ref{fig:setup20deIT}
correspond to the sparse and dense topology, respectively, of dataset
I, whereas Figs.~\ref{fig:setup20spNW} and \ref{fig:setup20deNW}
correspond to the sparse and dense topology, respectively, of dataset
N.

\begin{figure*}[tbp]
  \centering
  \subfloat[I dataset; sparse topology]{\label{fig:snrIs}\includegraphics[width=0.4\textwidth]{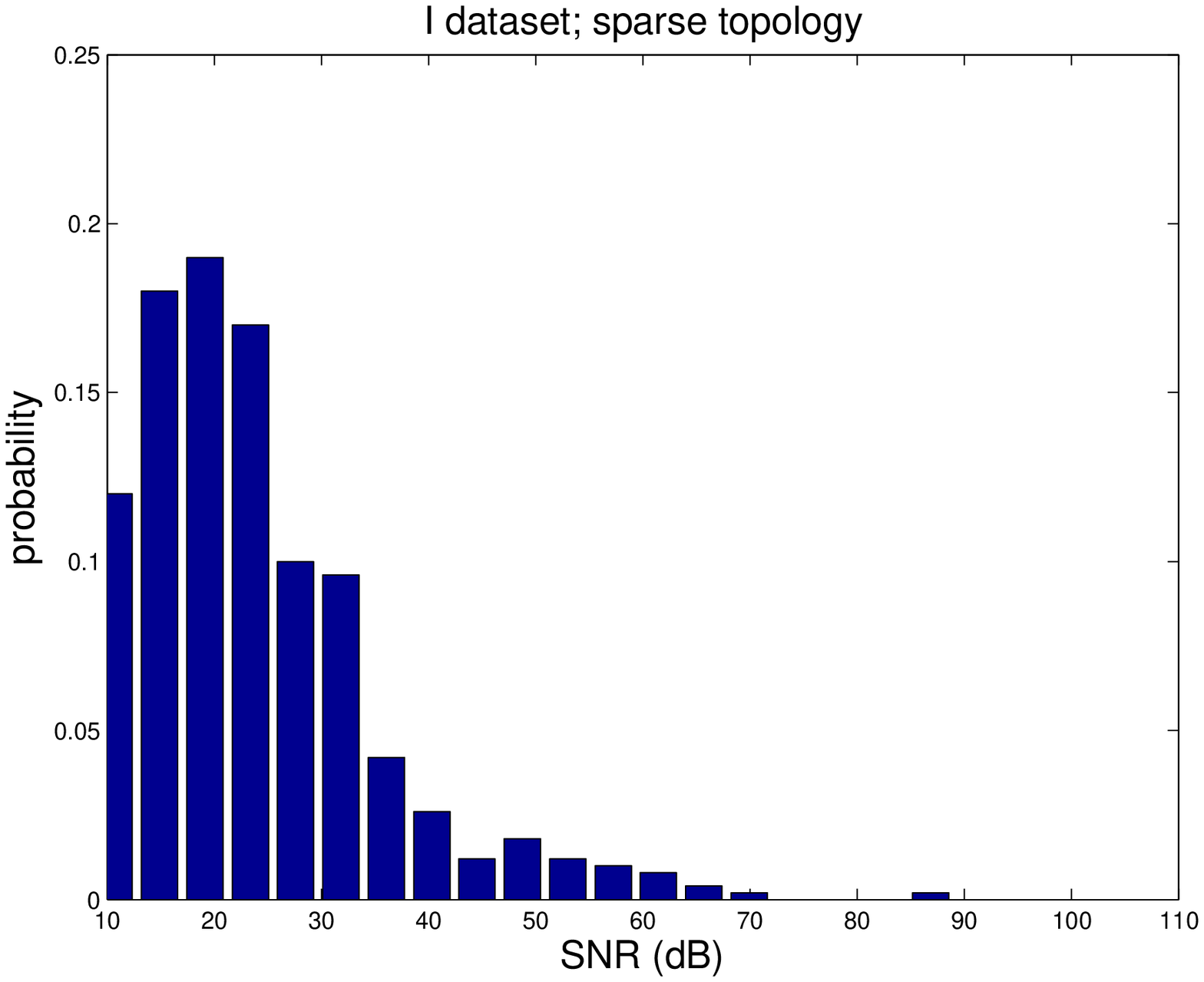}}
  \subfloat[N dataset; sparse topology]{\label{fig:snrNs}\includegraphics[width=0.4\textwidth]{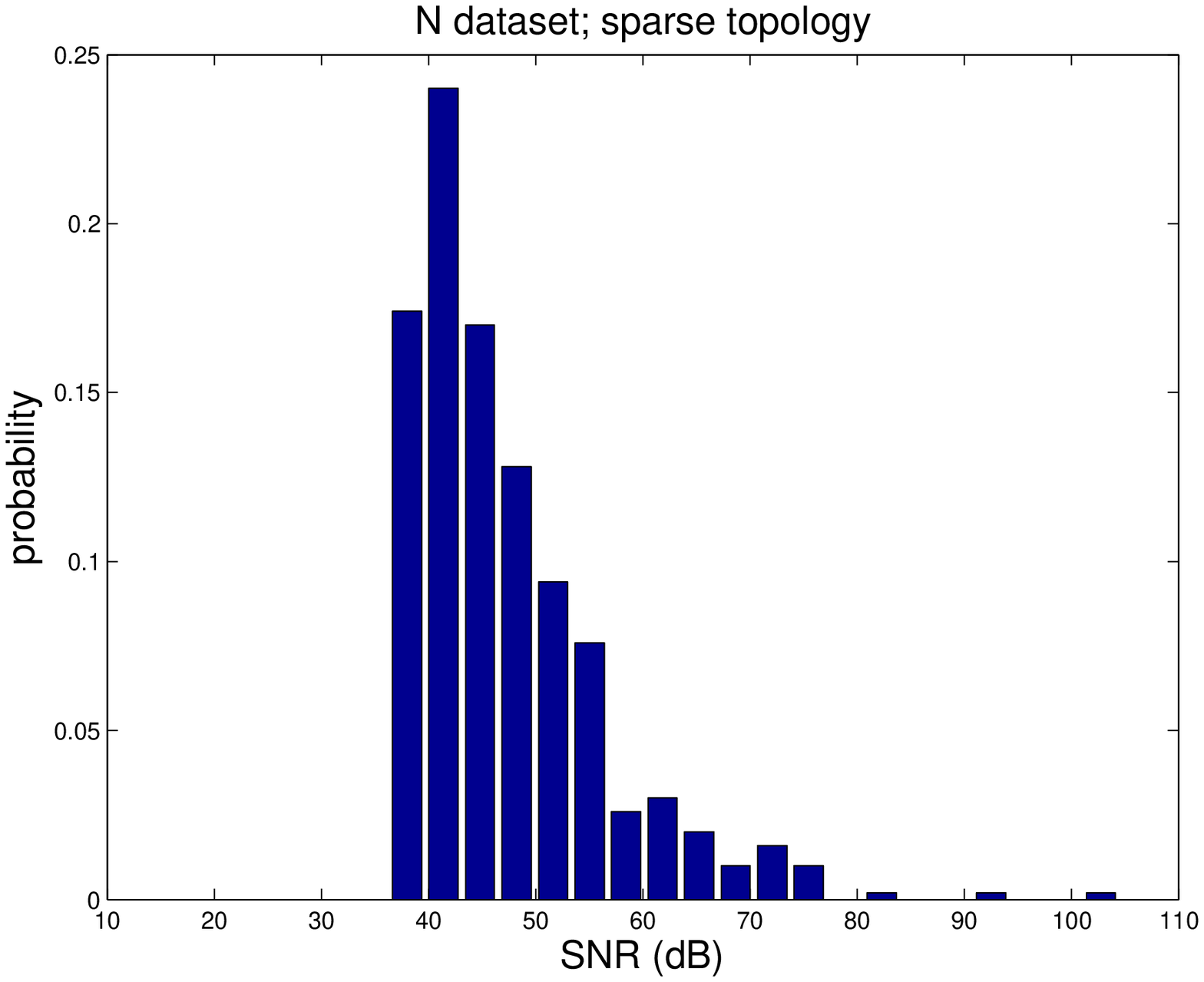}}\\
  \subfloat[I dataset; dense topology]{\label{fig:snrId}\includegraphics[width=0.4\textwidth]{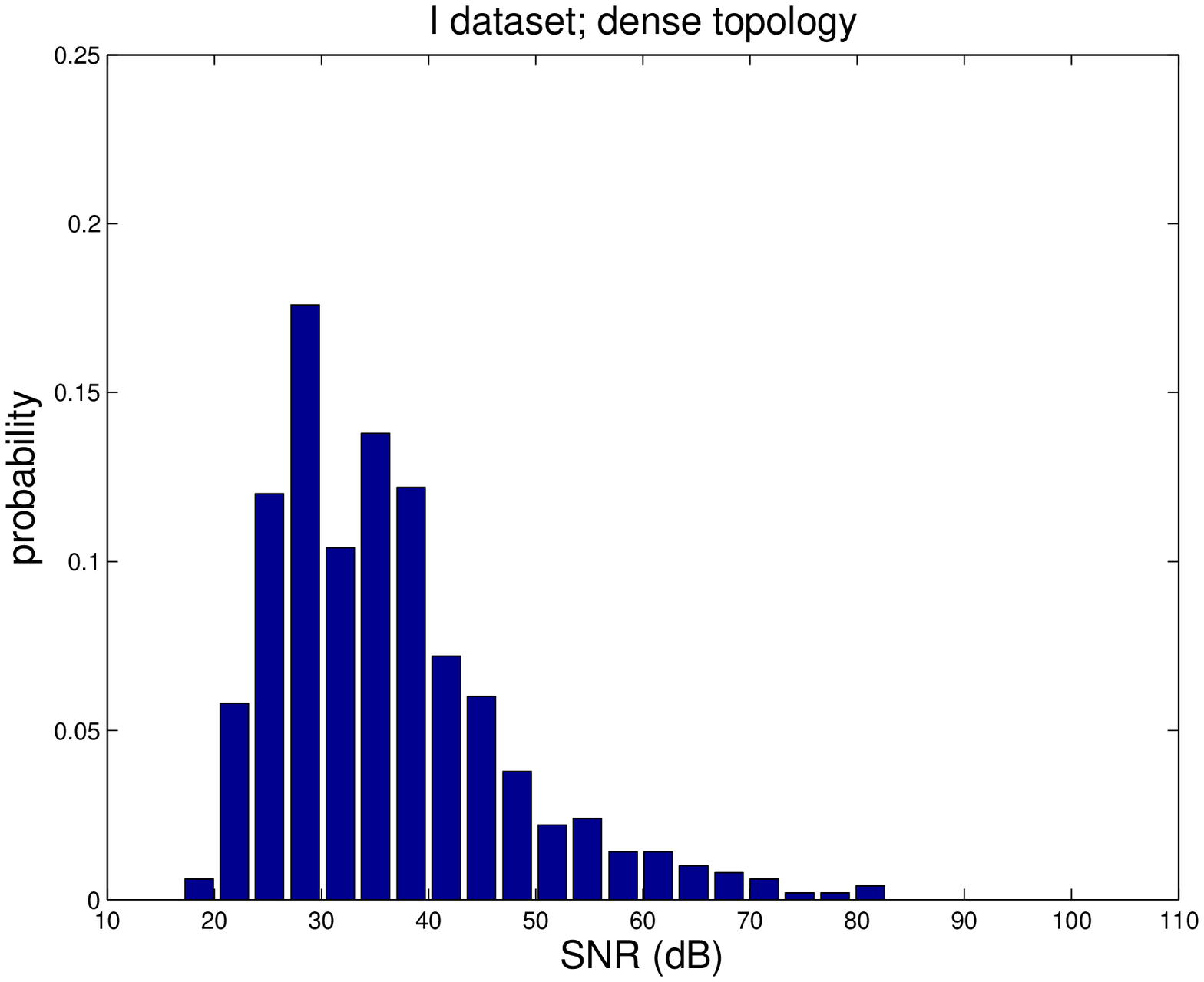}}
  \subfloat[N dataset; dense topology]{\label{fig:snrNd}\includegraphics[width=0.4\textwidth]{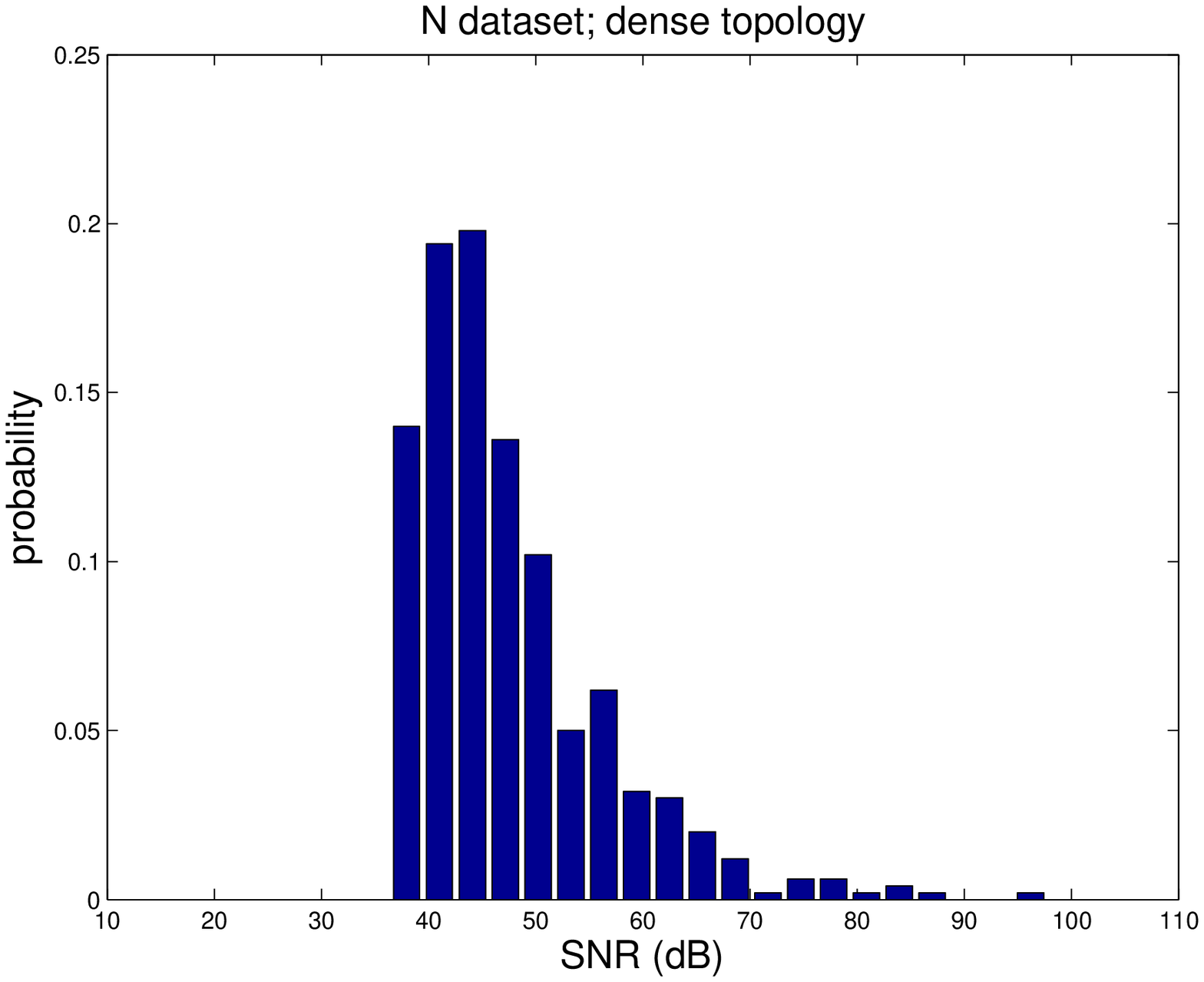}}
  \caption{Distribution of SNR for different datasets and densities.}
  \label{fig:SNR}
\end{figure*}

The input parameters are chosen to be common for all links; specifically, the transmit power $p_k$, $\forall k\in\K$, is set to $30$dBm, the noise power $\eta$ to $-100$dBm, and the channel gains $G_{mk}$ follow the geometric, distance-based, path loss model with an exponent of $4$. The major difference between the datasets is the distribution of the link lengths, which effectively determines the SNR distribution of the links. The input parameters yield minimum SNR approximately equal to $4$dB, $16$dB, and $32$dB for dataset sparse I, dense I, and N, respectively. The histograms in Fig.~\ref{fig:SNR} illustrate the SNR distribution of each dataset; as in Fig.~\ref{fig:instances}, left and right sub-figures are for dataset I and N, respectively, whereas upper and lower sub-figures are for sparse and dense topologies, respectively. For dataset I, the links in the sparse topology have on average lower SNR than in the dense topology; the mass of the SNR distribution is roughly for $10$--$40$dB in the sparse and for $20$--$50$dB in the dense topology. This is because in the sparse topology the test area is enlarged, allowing generation of longer links which have lower SNR values. On the contrary, the SNR distribution of dataset N is invariant to the network density; this is by construction, since the distribution of the link lengths is not affected by the size of the test area.

\begin{figure*}[tbp]
  \centering
  \subfloat[LA-SUD]{\label{fig:10Id-9-SUD}\includegraphics[width=0.4\textwidth]{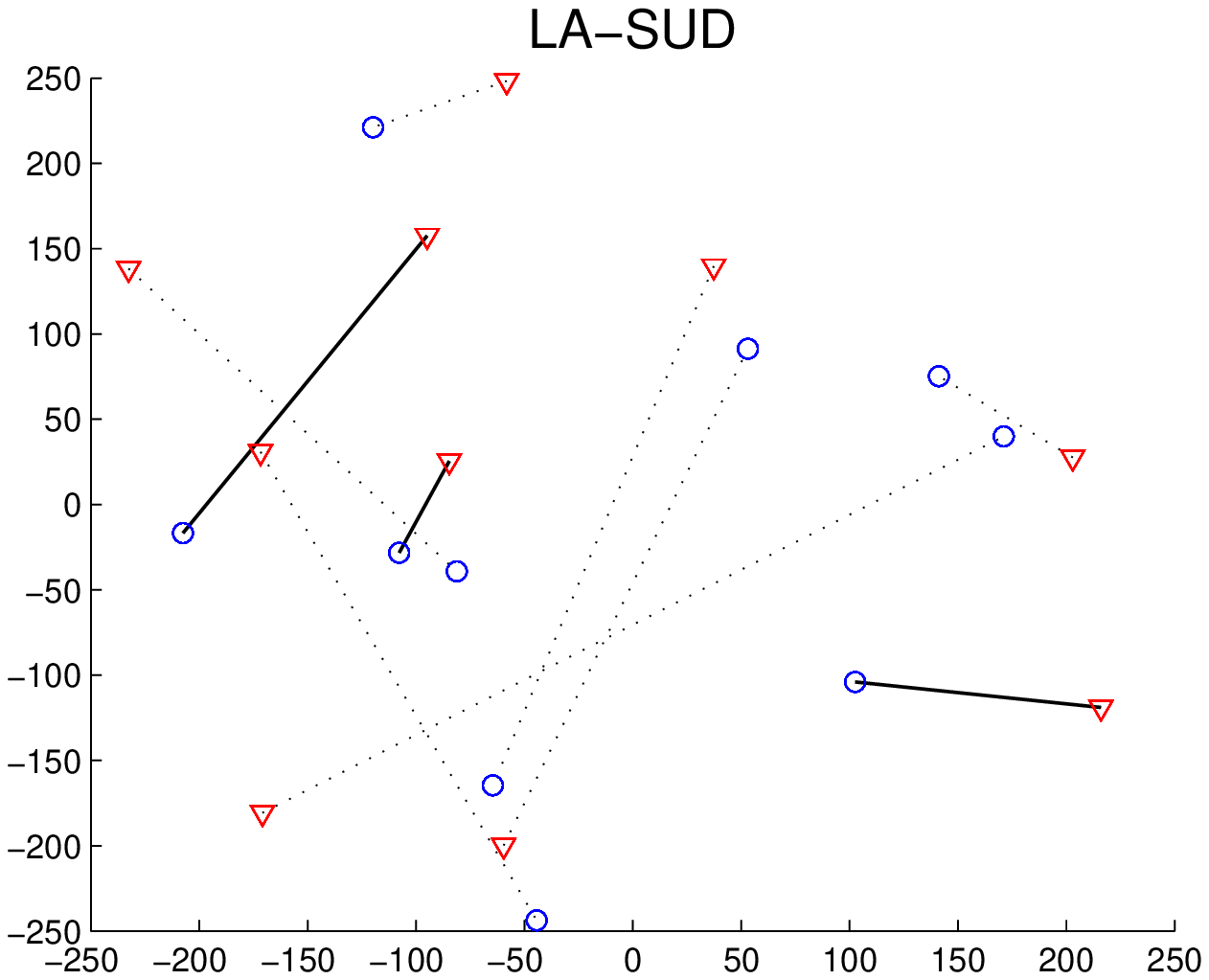}}
  \subfloat[LA-SLIC]{\label{fig:10Id-9-SLIC}\includegraphics[width=0.4\textwidth]{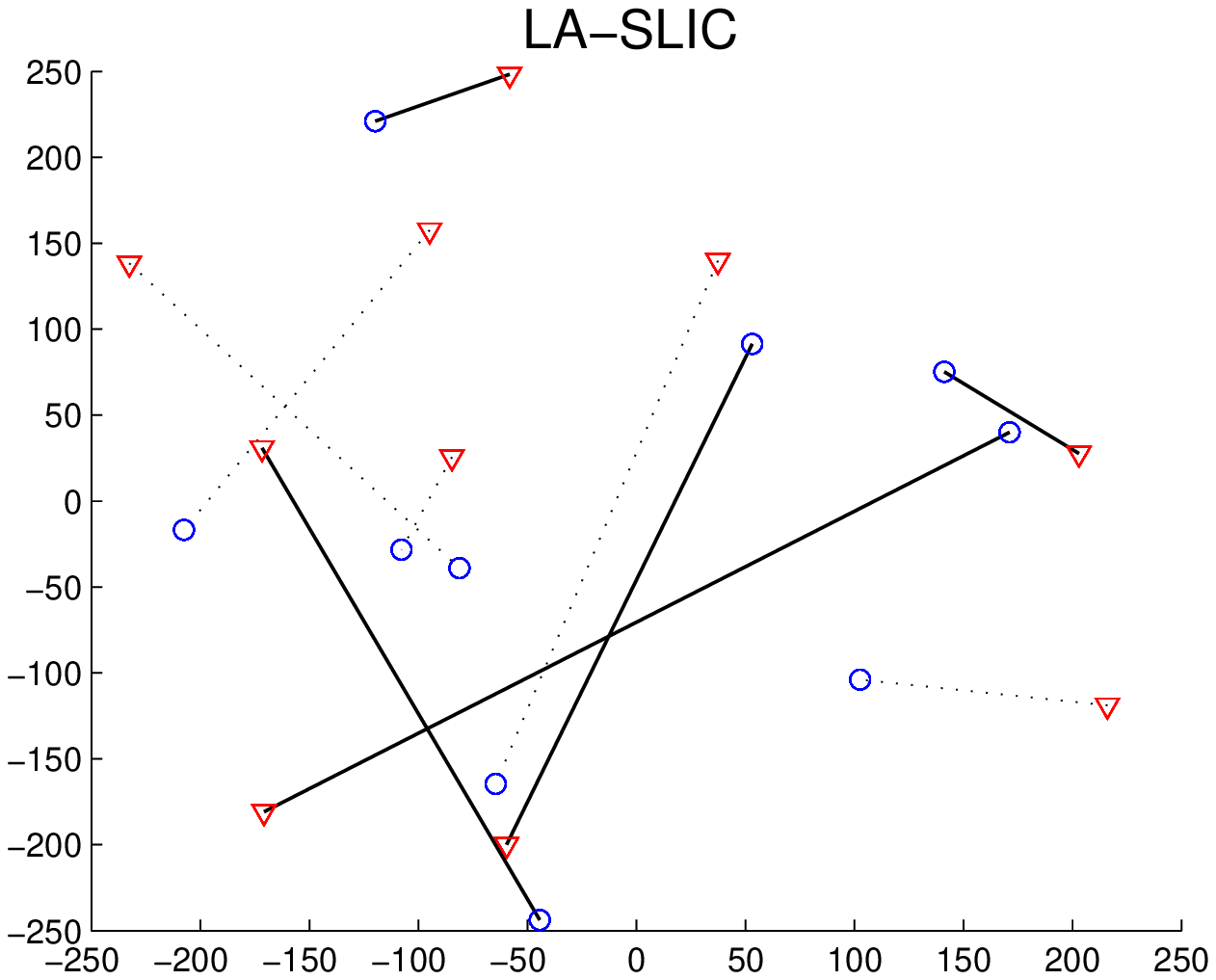}}\\
  \subfloat[LA-PIC]{\label{fig:10Id-9-PIC}\includegraphics[width=0.4\textwidth]{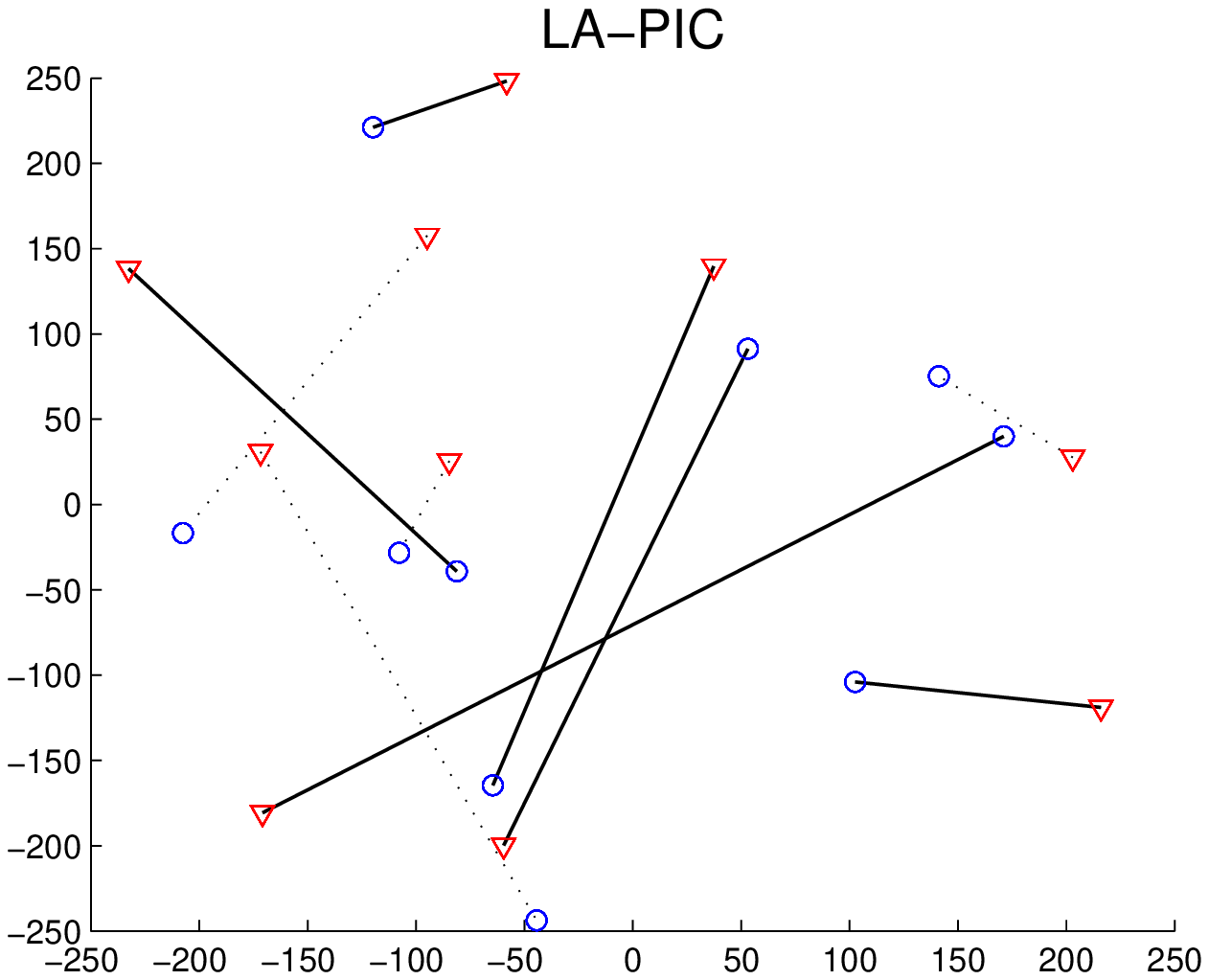}}
  \subfloat[LA-SIC]{\label{fig:10Id-9-SIC}\includegraphics[width=0.4\textwidth]{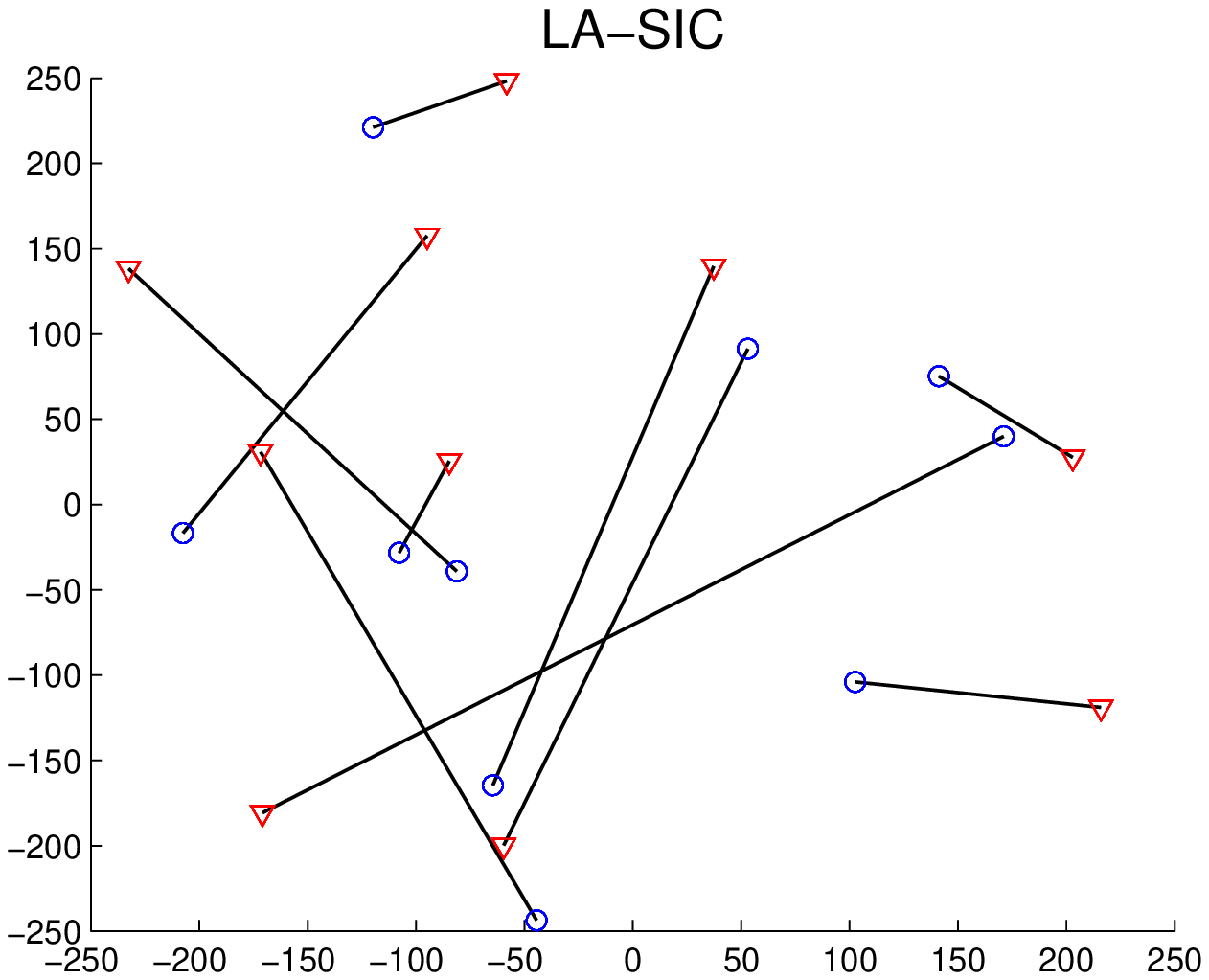}}
  \caption{Exemplary activation sets for different IC schemes; I dataset; dense topology; $10$ links; SINR $-9$dB.}
  \label{fig:activation}
\end{figure*}

For each dataset and network cardinality, $30$ instances are generated and the performance of LA with IC is assessed by two simulation studies. In the first study, all links are assumed to require for activation a common SINR threshold $\gamma_k = \gamma$, $\forall k\in\K$, taking values from $-9$dB up to $6$dB, and have equal activation weights, e.g., $w_k=1$, $\forall k\in\K$. The goal is to evaluate the performance gain due to single-link, parallel, and successive IC schemes on the LA problem over the baseline approach without IC. For this purpose, we implemented the formulations \eqref{eq:prelim:model}, \eqref{eq:SSIC_model}--\eqref{eq:SSIC_SLIC}, \eqref{eq:SSIC_model}, and \eqref{eq:ModelIc}, for LA-SUD, LA-SLIC, LA-PIC, and LA-SIC, respectively. Fig.~\ref{fig:activation} illustrates exemplary activation sets for an instance of a 10-link network, drawn from dataset dense I, when the SINR threshold is $-9$dB. It is evidenced that performance increases with problem sophistication: Figs. \eqref{fig:10Id-9-SUD}, \eqref{fig:10Id-9-SLIC}, \eqref{fig:10Id-9-PIC}, and \eqref{fig:10Id-9-SIC}, show that LA-SUD, LA-SLIC, LA-PIC, and LA-SIC activate $3$, $5$, $6$, and $10$ links, respectively.

The optimal solutions are found by an off-the-shelf solver, implementing standard techniques such as branch-and-bound and cutting planes \cite{Bert97}.
The simulations were performed on a server with a quad-core AMD Opteron processor at 2.6 GHz and 7 GB of RAM. The ILP formulations were implemented in AMPL 10.1 using the Gurobi Optimizer ver. 3.0.
Regarding the computational complexity of the proposed ILP formulations for IC, an empirical measure is the running time of the solution process.
We have observed that it is not an obstacle for practical instance sizes.

\begin{figure*}[tbp]
\centering
\subfloat[I dataset; sparse topology] {\label{fig:sinr-6Is}\includegraphics[width=0.4\textwidth]{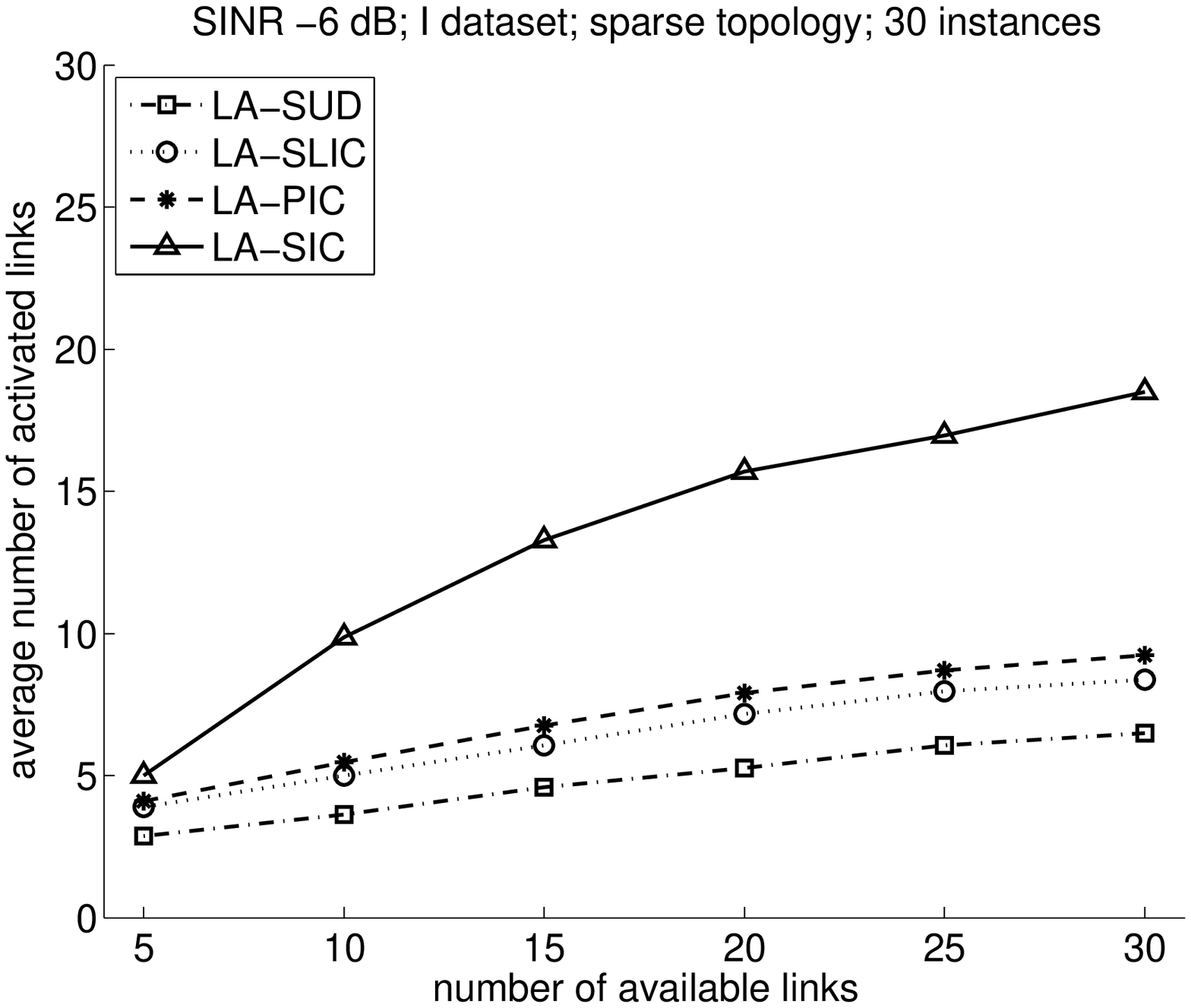}}
\subfloat[N dataset; sparse topology] {\label{fig:sinr-6Ns}\includegraphics[width=0.4\textwidth]{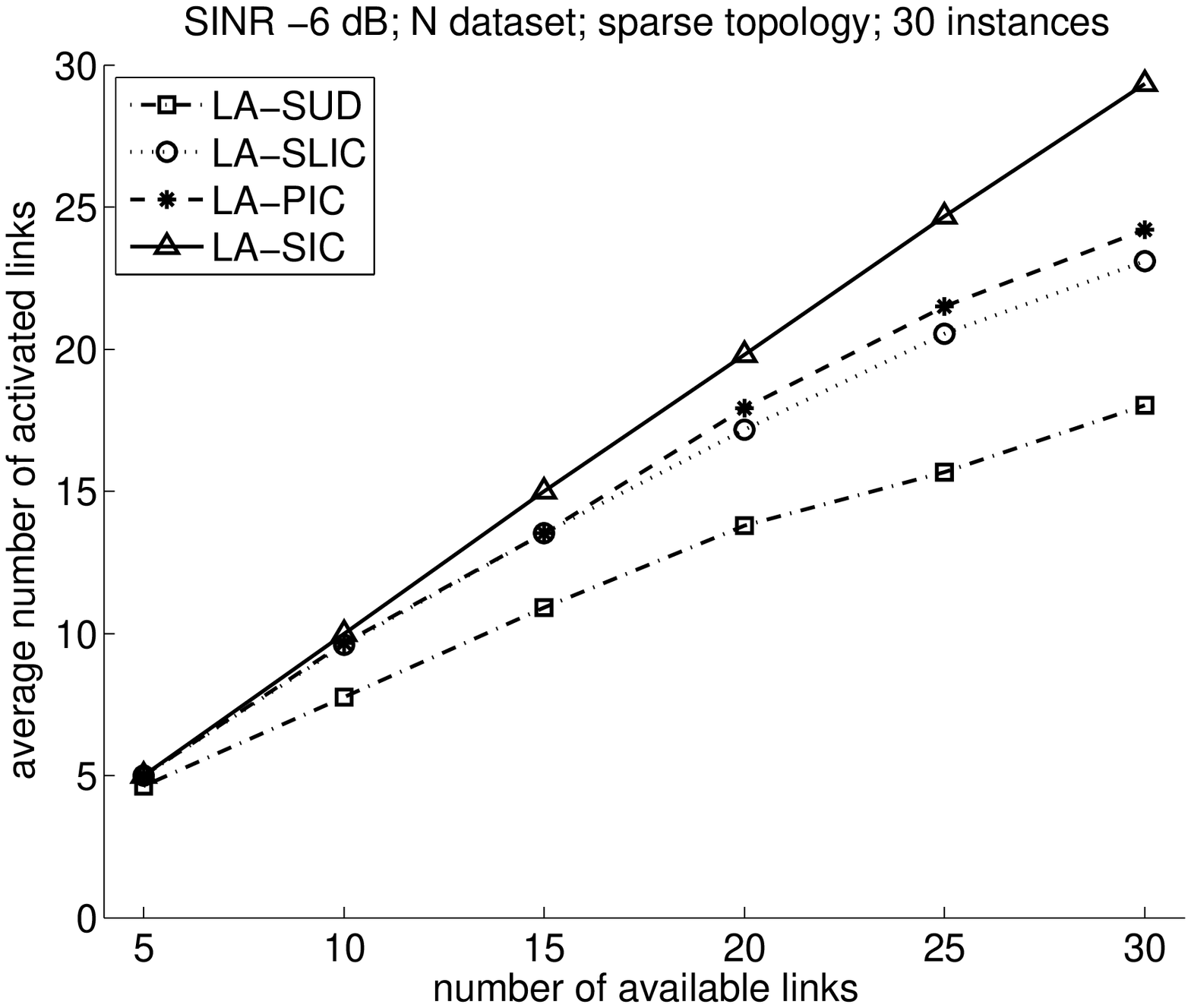}}\\
\subfloat[I dataset; dense topology] {\label{fig:sinr-6Id}\includegraphics[width=0.4\textwidth]{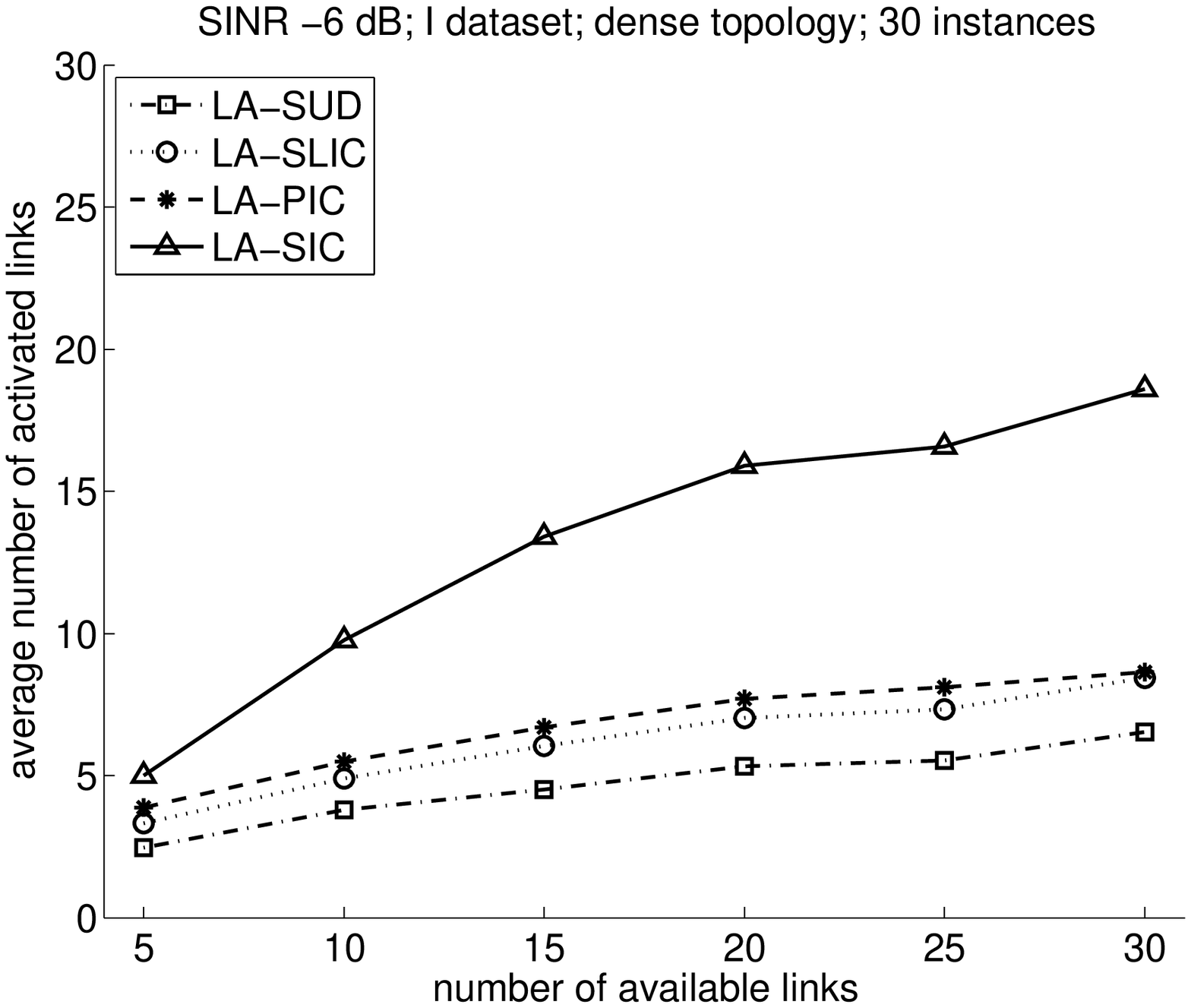}}
\subfloat[N dataset; dense topology] {\label{fig:sinr-6Nd}\includegraphics[width=0.4\textwidth]{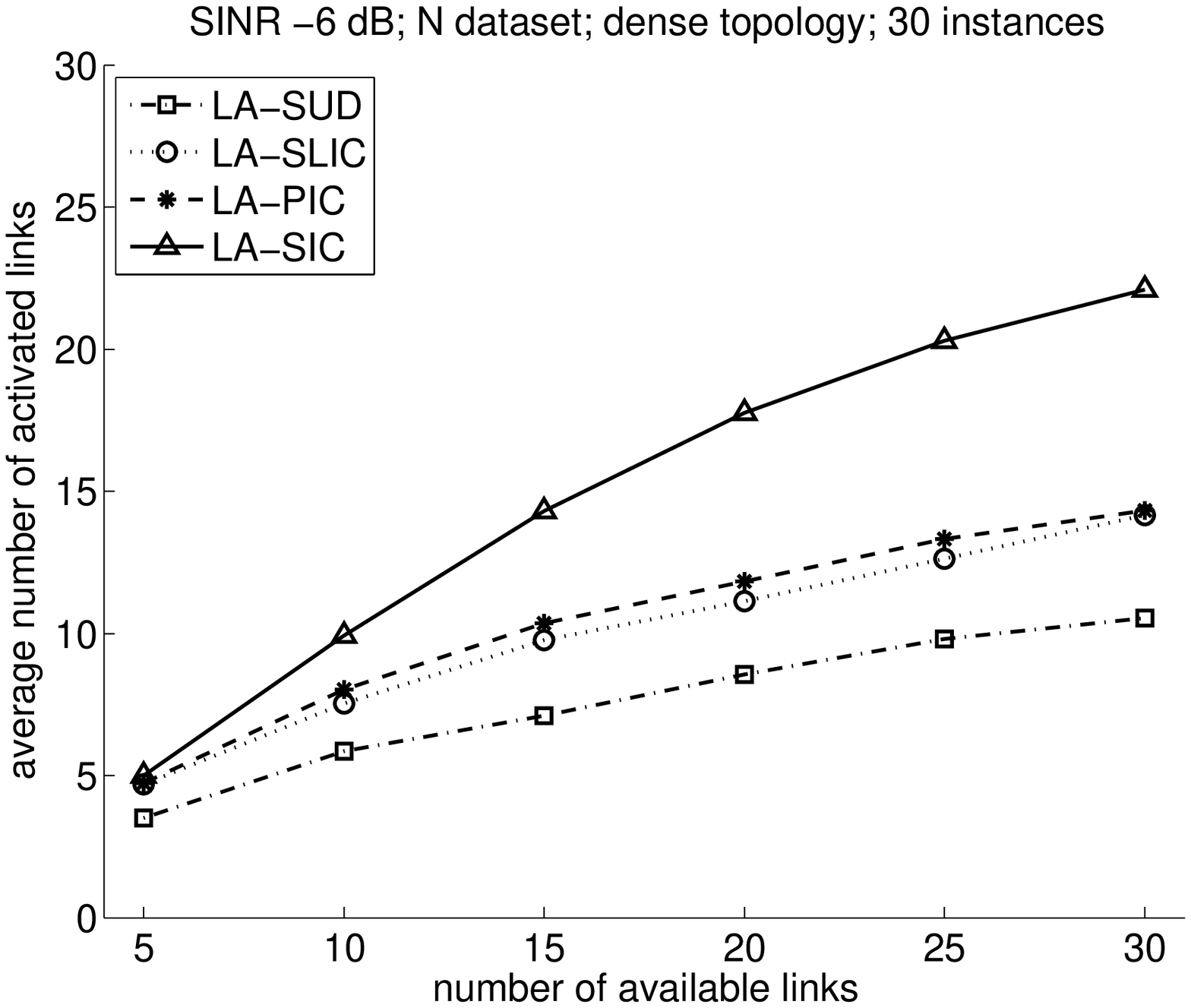}}
\caption{Average number of activated links versus network size for SINR threshold $-6$dB.}
\label{fig:SINR-6}
\end{figure*}

In the following a selection of the simulation results is presented. Fig.~\ref{fig:SINR-6} shows the average, over $30$ instances, number of activated links versus the total number of links in the network, achieved by all versions of the LA problem when the SINR threshold is $-6$dB. The results in the four sub-figures correspond to the datasets exemplified in Fig.~\ref{fig:instances}. The major observation is that all LA schemes with IC clearly outperform LA-SUD and in particular LA-SIC yields impressive performance. Comparing Figs.~\ref{fig:sinr-6Is} and~\ref{fig:sinr-6Id}, it is concurred that the results for dataset I are density invariant. As the number of links in the network increases, the performance of LA-SUD improves, due to the diversity, almost linearly but with very small slope. LA-SIC though improves significantly, activating two to three times more links than the baseline. When the network has up to about $15$ links, nearly all of them are activated with LA-SIC. On the other hand, LA-PIC has a consistent absolute gain over LA-SUD, activating one to two links more. Furthermore, LA-SLIC has almost as good performance with LA-PIC, i.e., it captures most of the gain due to single-stage IC. Fig.~\ref{fig:sinr-6Nd} shows that the LA schemes have similar performance in dataset dense N as in dataset I. Fig.~\ref{fig:sinr-6Ns} shows that LA is easier for dataset sparse N, even without IC. The curves of all schemes linearly increase with network cardinality, but with IC the slopes are higher, so that the absolute gains, differences from the baseline, broaden. Maximum gains are for $30$ links, where LA-SUD, LA-SLIC, LA-PIC, and LA-SIC activate about $18$, $22$, $23$, and $30$ links, respectively. For the tested network cardinalities, LA-SIC achieves the ultimate performance activating all links.

\begin{figure*}[tbp]
\centering
\subfloat[I dataset; sparse topology] {\label{fig:sinr3Is}\includegraphics[width=0.4\textwidth]{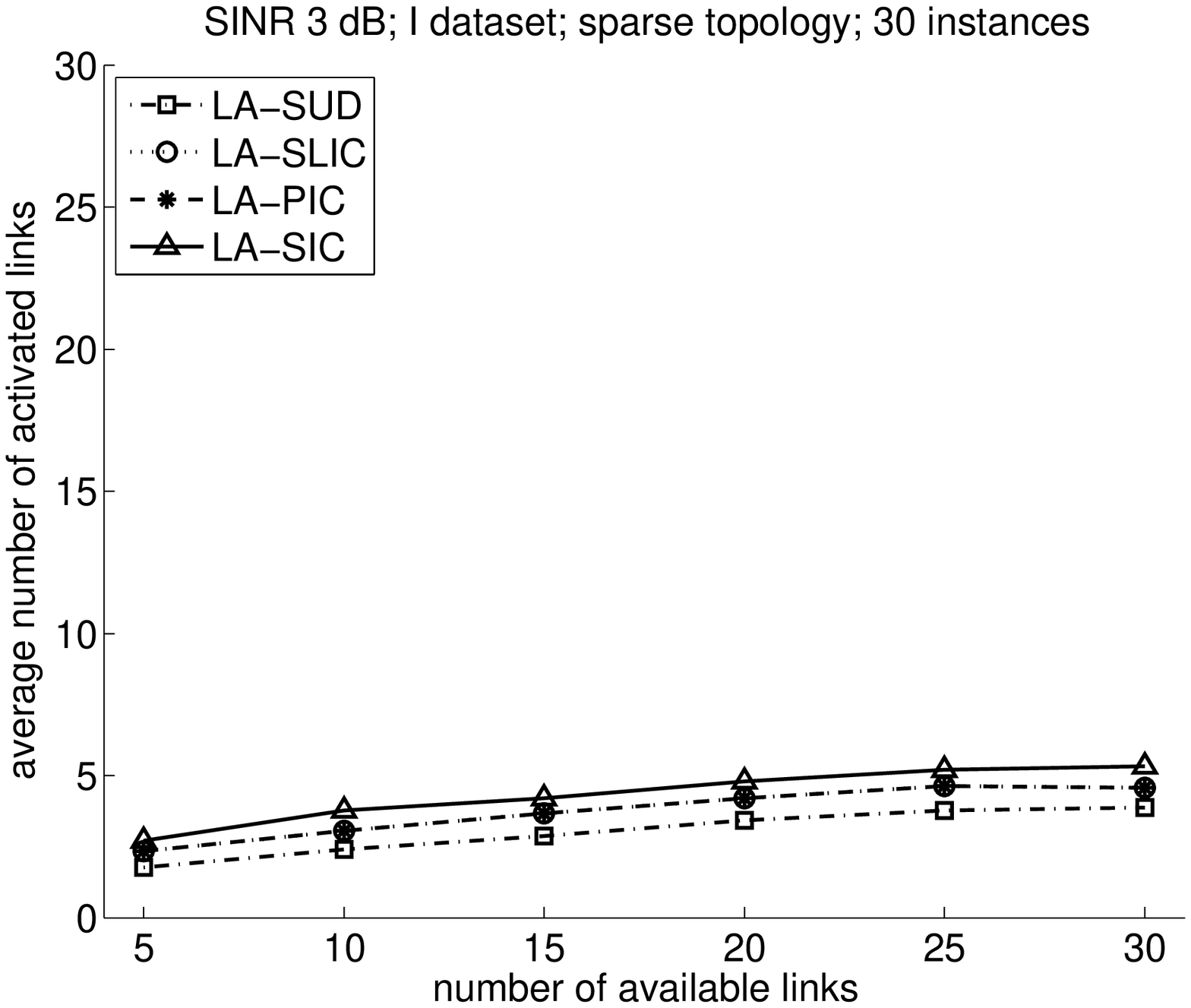}}
\subfloat[N dataset; sparse topology] {\label{fig:sinr3Ns}\includegraphics[width=0.4\textwidth]{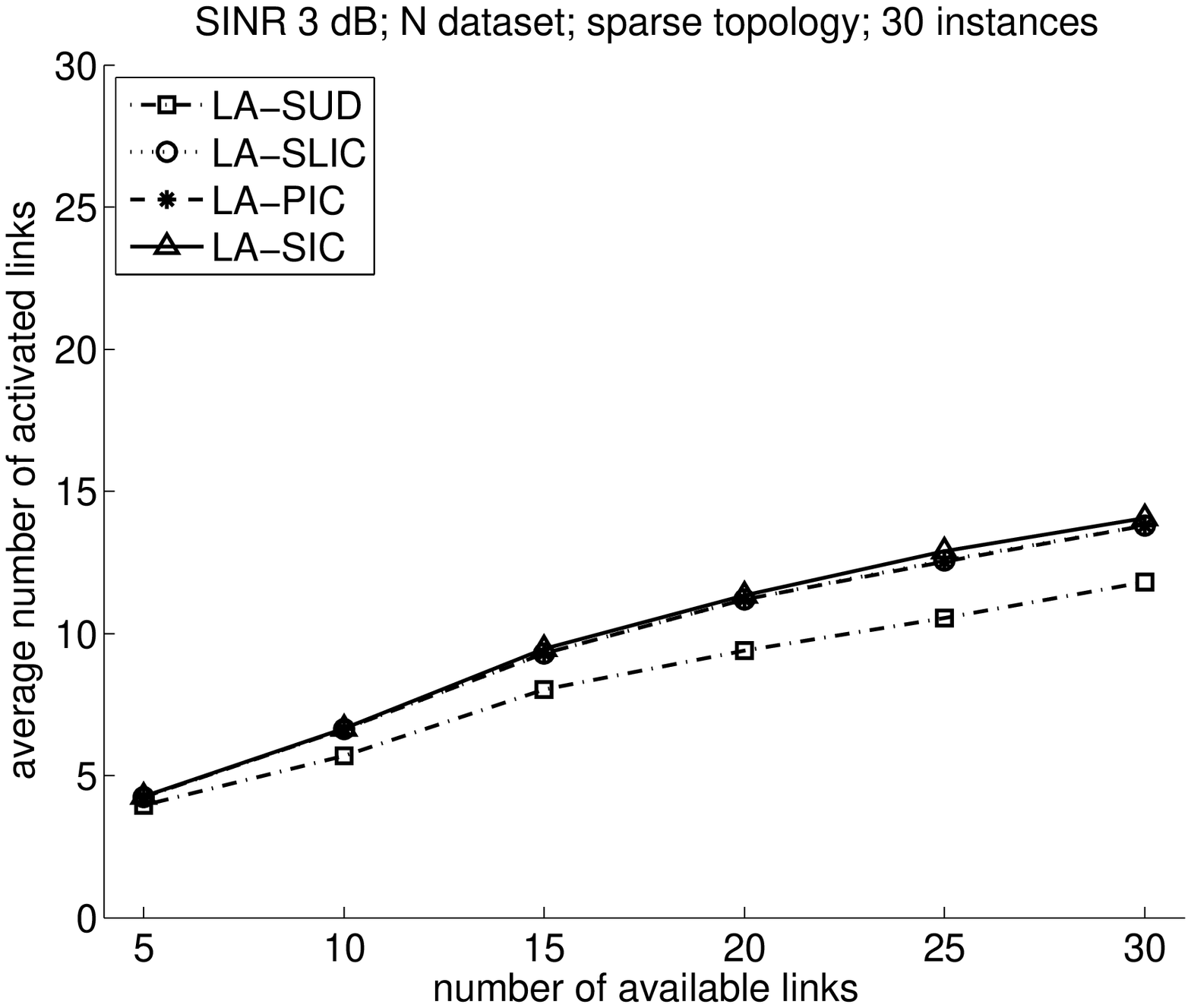}}\\
\caption{Average number of activated links versus network size for SINR threshold $3$dB.}
\label{fig:SINR3}
\end{figure*}

As seen in Fig.~\ref{fig:SINR-6}, the performance gains due to IC are very significant when the SINR threshold for activation is low. However, for high SINR thresholds, the gains are less prominent. For example, Fig.~\ref{fig:SINR3} shows the performance of the LA schemes when $\gamma$ is set to $3$dB. Figs.~\ref{fig:sinr3Is} and~\ref{fig:sinr3Ns} are for the sparse datasets I and N, respectively; for the dense topologies the results are similar to Fig.~\ref{fig:sinr3Is}. It is evidenced that IC schemes activate one to two links more than the baseline and that most of this gain can be achieved with single-stage IC.

\begin{figure*}[tbp]
\centering
\subfloat[I dataset; sparse topology] {\label{fig:size30Is}\includegraphics[width=0.4\textwidth]{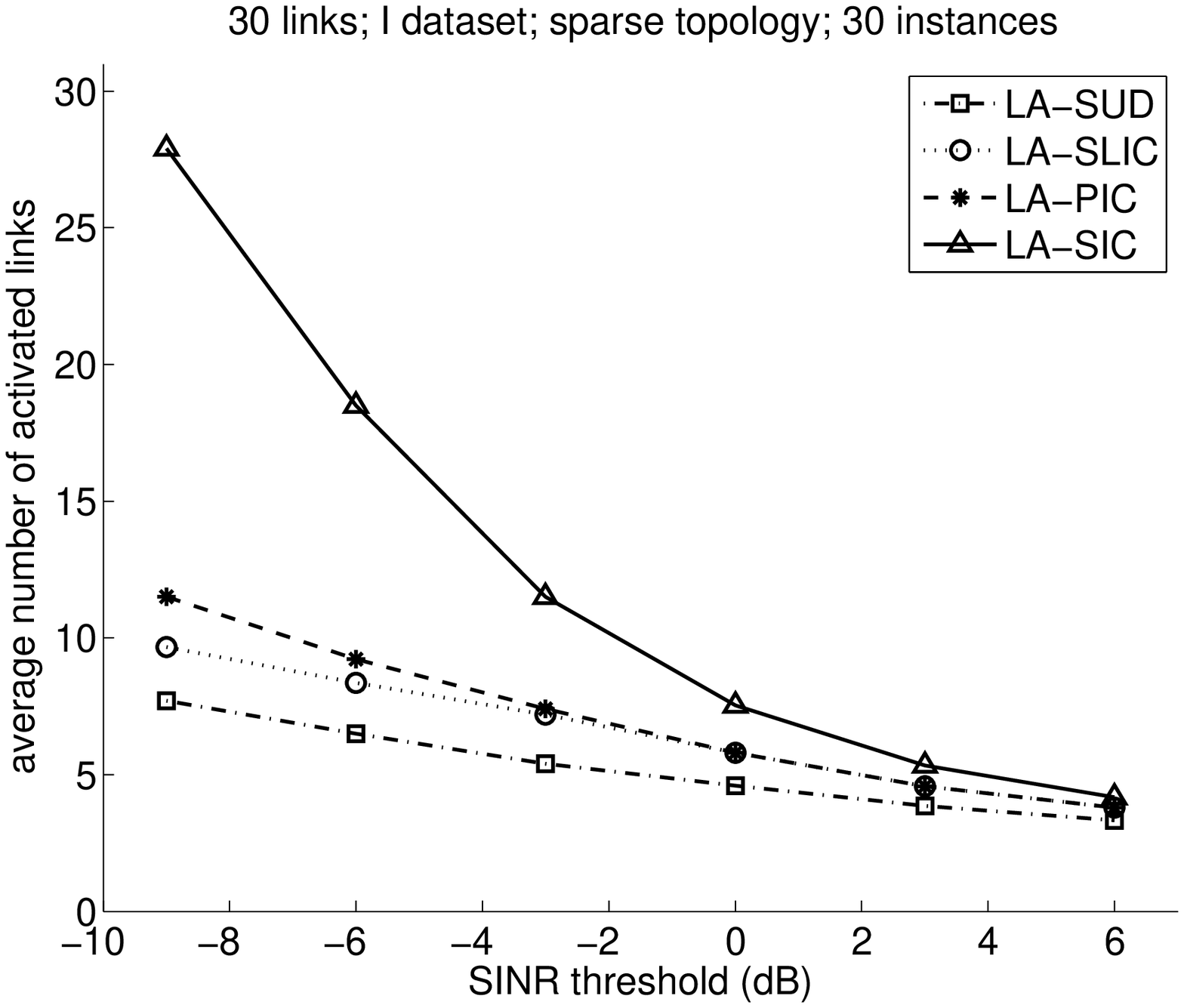}}
\subfloat[N dataset; sparse topology] {\label{fig:size30Ns}\includegraphics[width=0.4\textwidth]{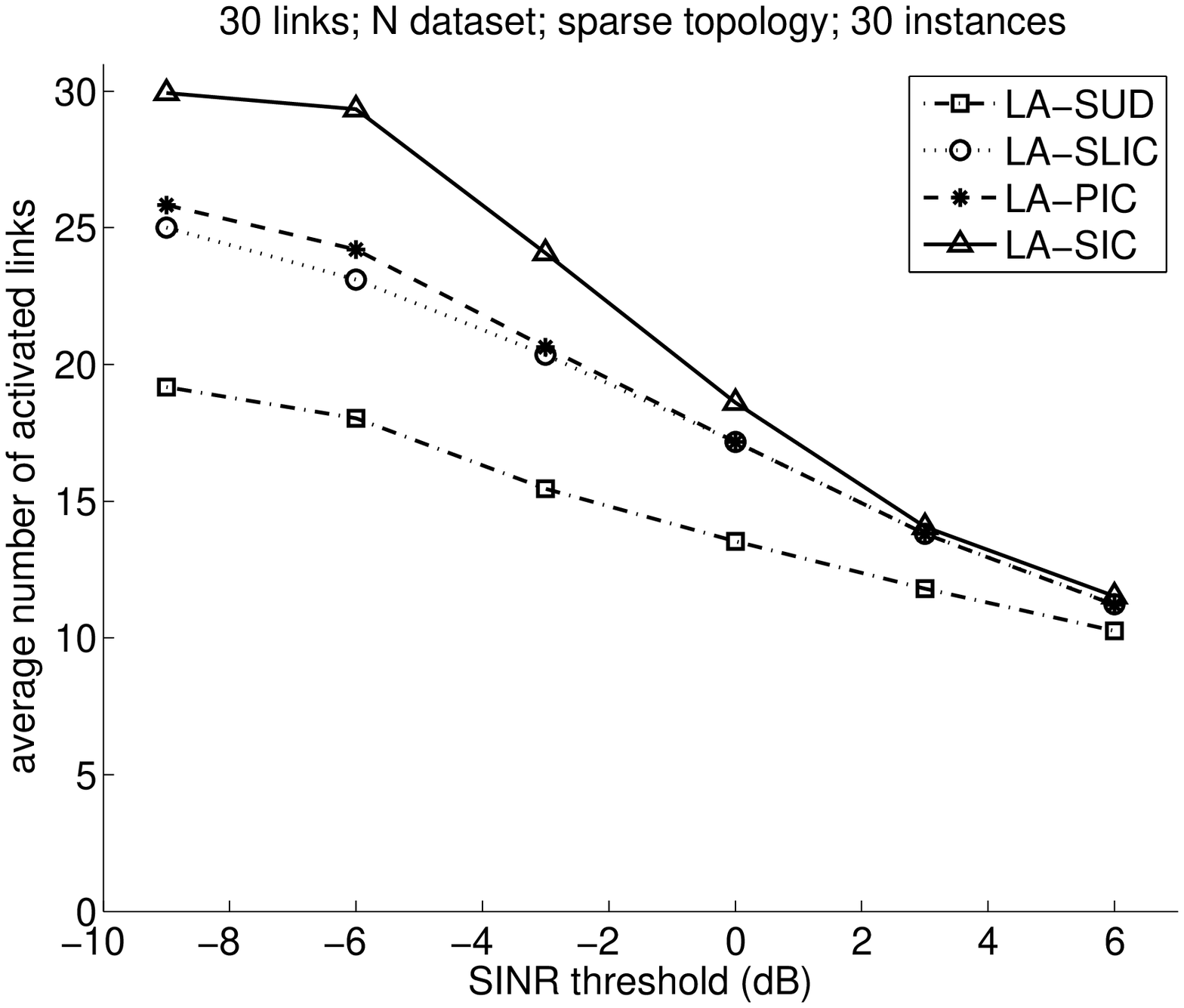}}\\
\subfloat[I dataset; dense topology] {\label{fig:size30Id}\includegraphics[width=0.4\textwidth]{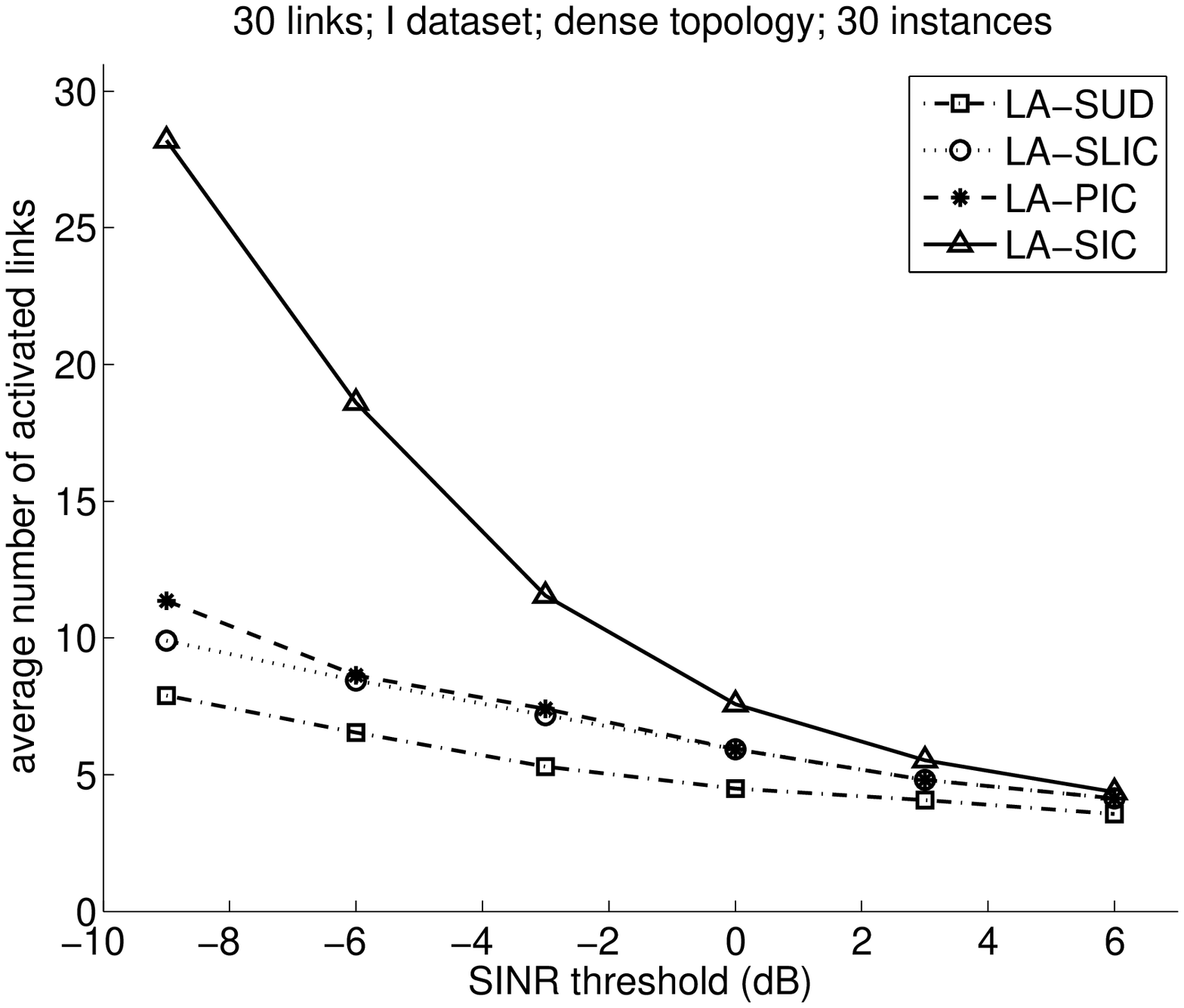}}
\subfloat[N dataset; dense topology] {\label{fig:size30Nd}\includegraphics[width=0.4\textwidth]{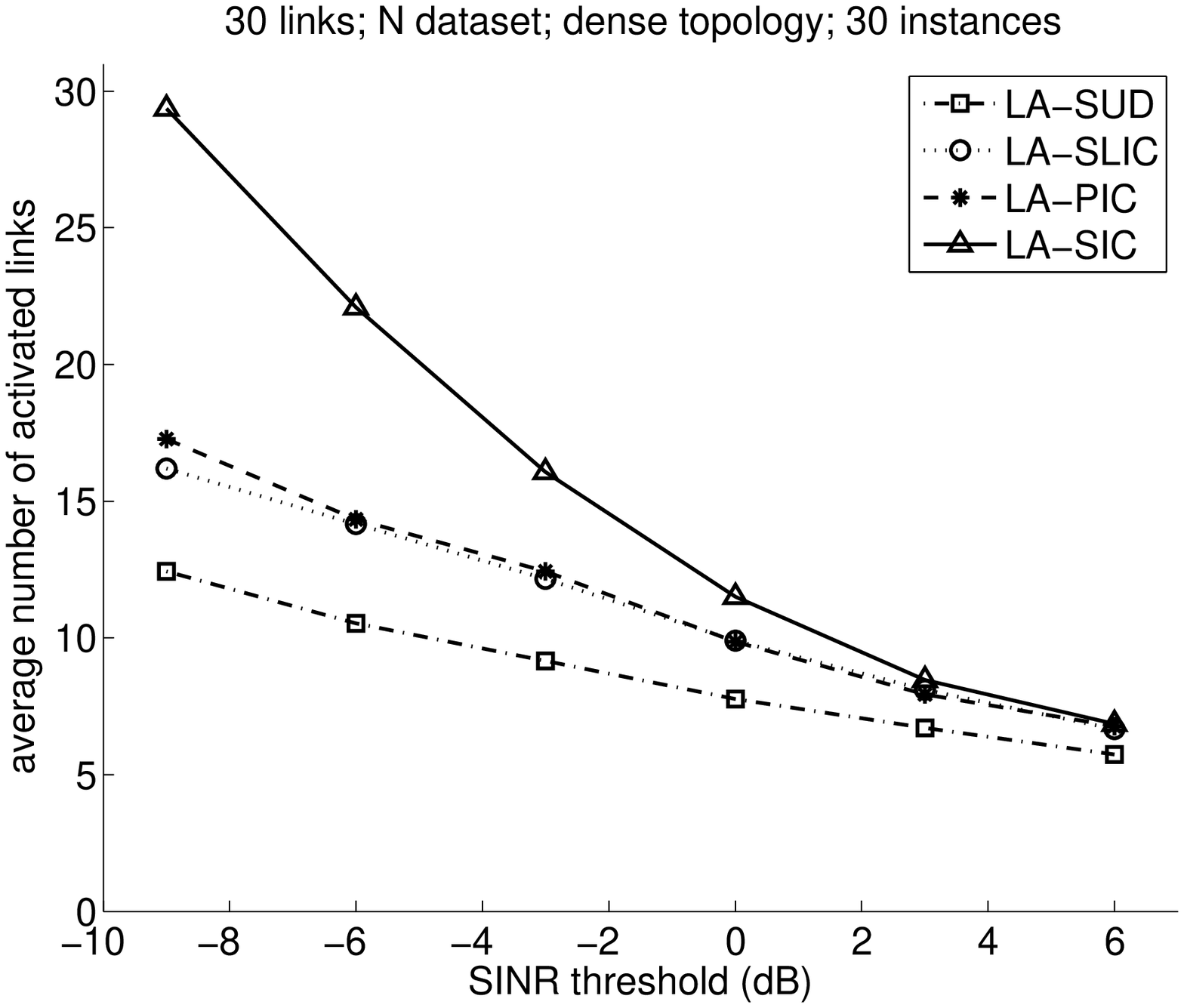}}
\caption{Average number of activated links versus SINR threshold for network of $30$ links.}
\label{fig:size30}
\end{figure*}

The fact that the IC gains diminish with increasing the SINR threshold is clearly illustrated in Fig.~\ref{fig:size30}, which compares, for networks of 30 links, the average performance of all LA schemes for various SINR thresholds. The relative gain of SIC is more prominent in the case of dataset I, which is more challenging for the baseline problem. For dataset I, when the SINR threshold is low, around $-9$dB, SIC activates nearly all links, whereas SUD activates less than a third of them. For sparse and dense dataset N, SIC activates effectively all links when the SINR threshold is lower than $-6$dB and $-9$dB, respectively, whereas SUD activates less than two thirds and less than half of them, respectively. For mid-range SINR thresholds, up to about $3$dB, SIC has an exponentially decreasing performance, but nevertheless still significantly outperforms SUD. On the other hand, for SINR thresholds up to about $0$dB, PIC yields a relatively constant performance improvement of roughly two to five links, depending on the dataset. PIC is effectively equivalent to its simpler counterpart SLIC, for SINR higher than $-6$dB. The performance of all IC schemes converges for SINR thresholds higher than 3dB. The interpretation is that if IC is possible, it is more likely that it will be restricted to a single link. For very high SINR thresholds, it becomes rarely possible to perform IC.

\begin{figure*}[tbp]
\centering
\subfloat[I dataset; sparse topology] {\label{fig:indIs}\includegraphics[width=0.4\textwidth]{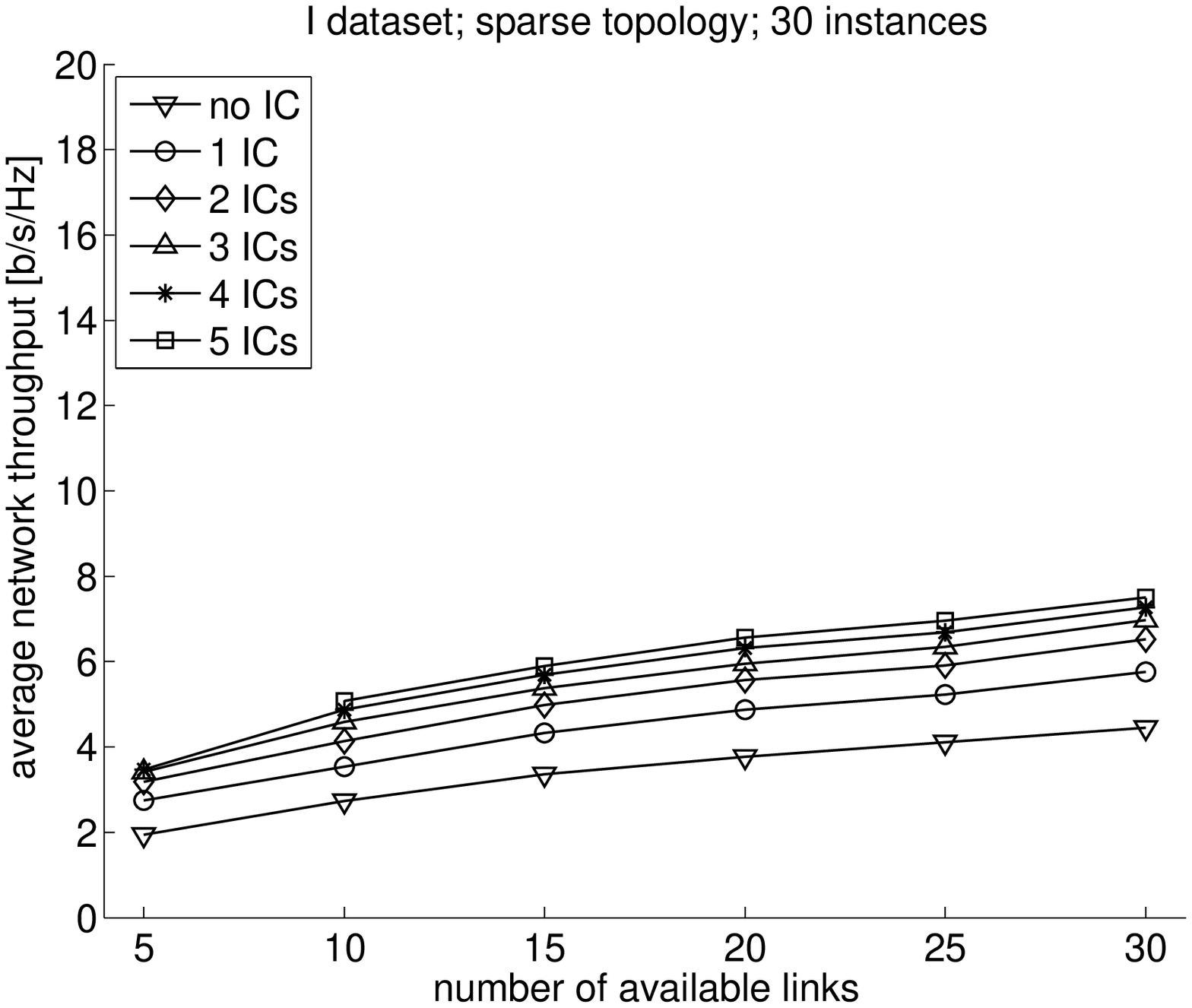}}
\subfloat[N dataset; sparse topology] {\label{fig:indNs}\includegraphics[width=0.4\textwidth]{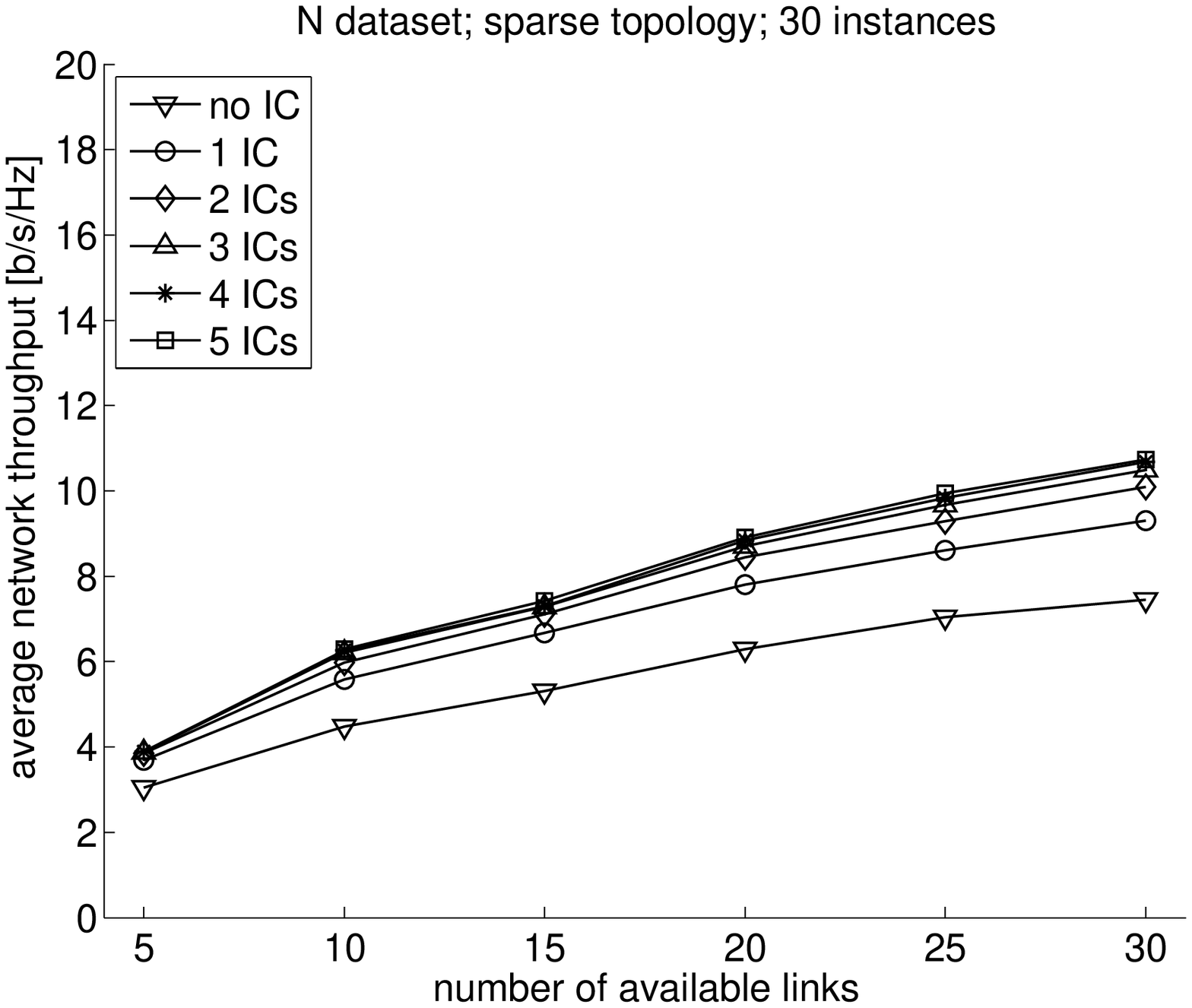}}\\
\subfloat[I dataset; dense topology] {\label{fig:indId}\includegraphics[width=0.4\textwidth]{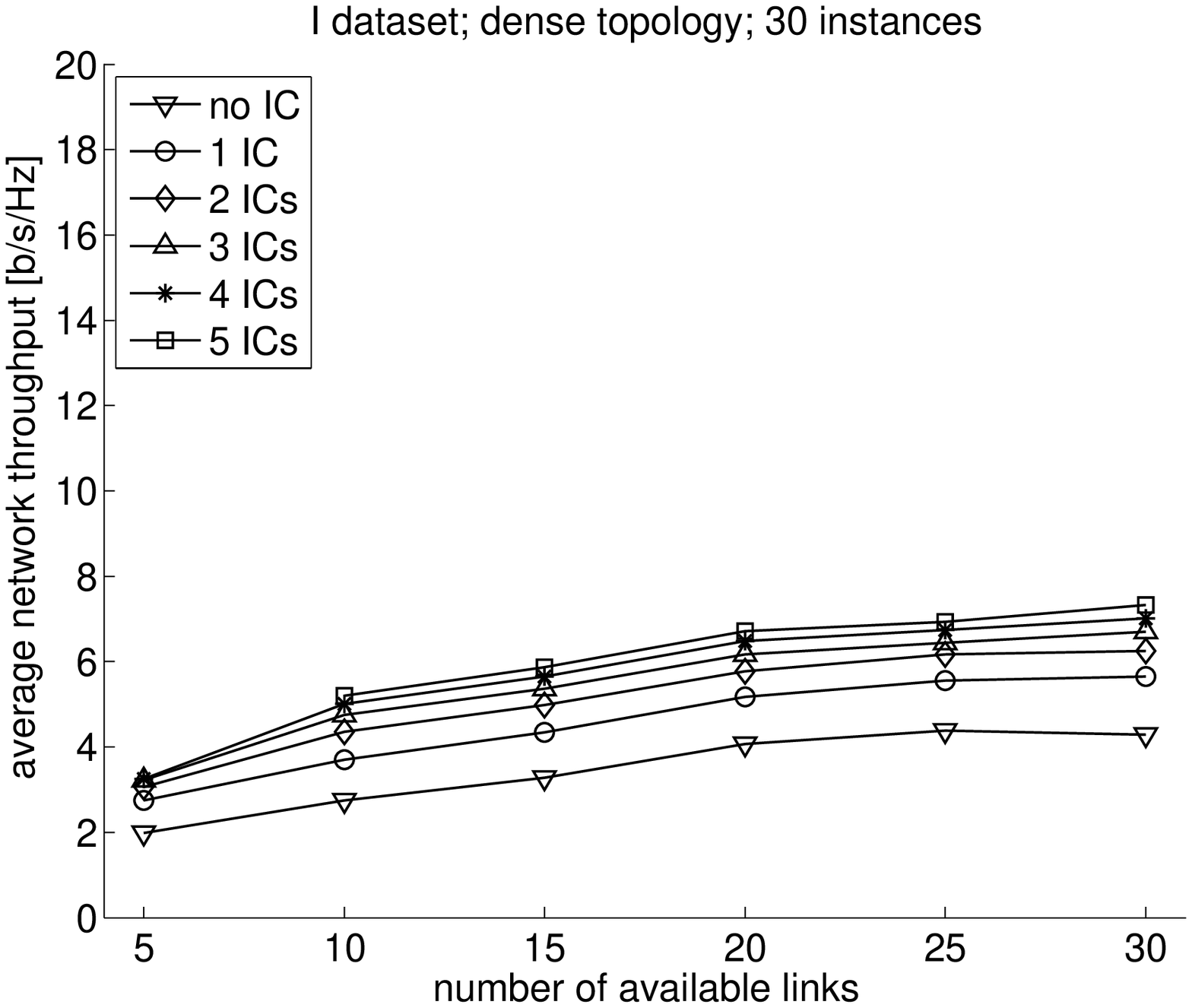}}
\subfloat[N dataset; dense topology] {\label{fig:indNd}\includegraphics[width=0.4\textwidth]{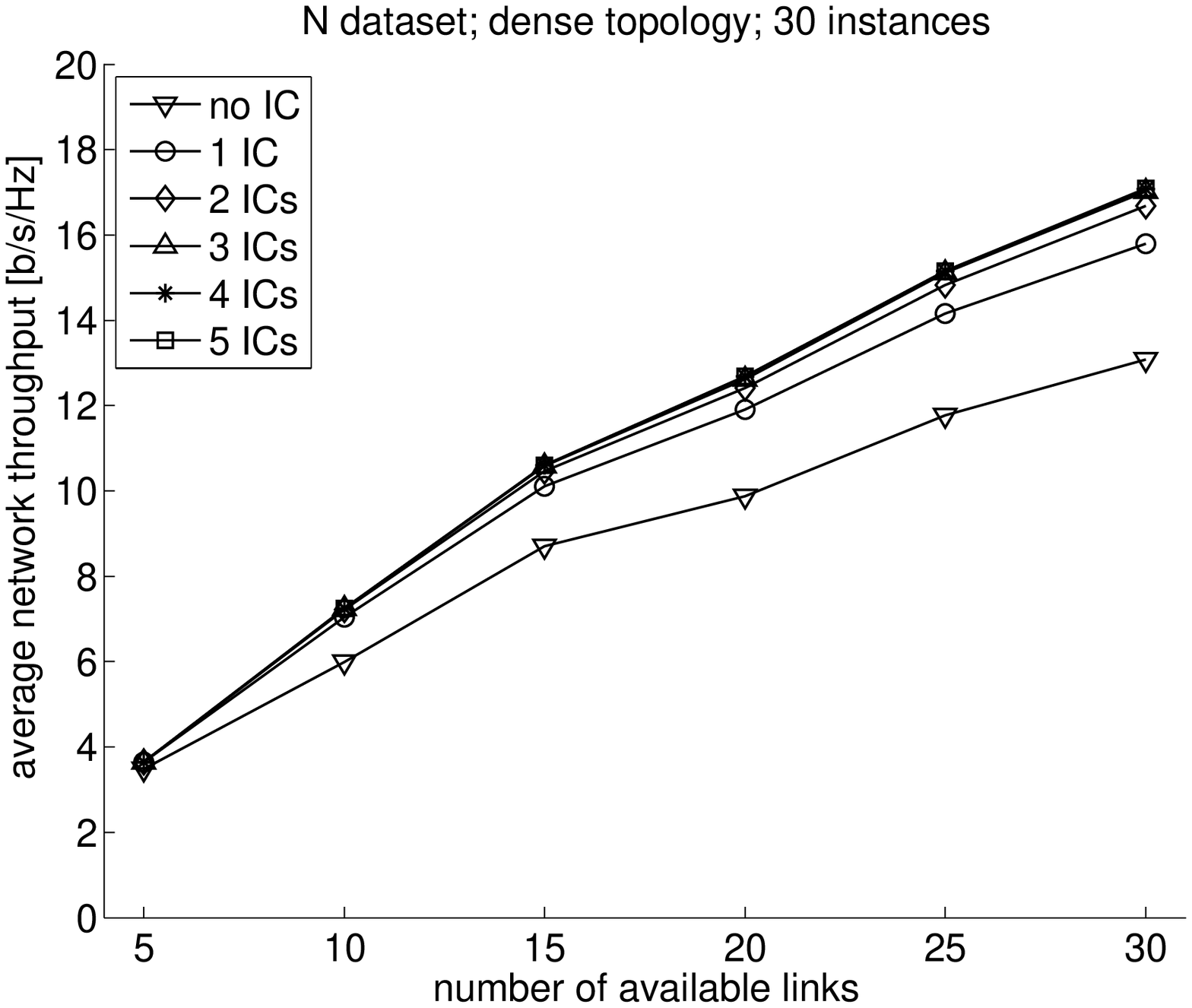}}
\caption{Average network throughput versus network cardinality for various IC schemes and SINR thresholds randomly chosen from $\{-6,-3,3\}$dB.}
\label{fig:individual}
\end{figure*}

In the second simulation setup, the performance of the general LA-SIC problem, under individual SINR thresholds, is evaluated. The SINR threshold $\gamma_k$ for each link is taking, with equal probability, one of the values in the set $\{-6,-3,3\}$dB and the activation weights $w_k$ are set equal to the data rates, in bits per second per Hertz, corresponding to the respective SINR thresholds. The formulation \eqref{eq:FinalModel} is implemented varying the maximum number of cancellation stages $T_k = T$, $\forall k\in\K$, from $0$, corresponding to the baseline case without IC, up to $5$. Fig.~\ref{fig:individual} shows the average, over 30 instances, throughput of all activated links versus the network cardinality, for all the datasets. For dataset I, the network throughput is almost doubled with IC; roughly half of this increase is achieved by the first cancellation stage and most of the rest by the next two to three stages. For dataset N, it is seen that the first cancellation stage yields a significant gain of about $2$ b/s/Hz and that it only pays off to have more than two cancellation stages for large and sparse networks.

\section{Conclusions}
\label{sec:conclusion}

\vspace*{-1mm}
In this paper, we have addressed the problem of optimal concurrent link activation in wireless systems with interference cancellation.
We have proved the NP-hardness of this problem and developed integer linear programming formulations that can be used to approach the exact optimum for parallel and successive interference cancellation.
Using these formulations, we have performed numerical experiments to quantitatively evaluate the gain due to interference cancellation.
The simulation results indicate that for low to medium SINR thresholds, interference cancellation delivers a significant performance improvement.
In particular, the optimal SIC scheme can double or even triple the number of activated links.
Moreover, node density may also affects performance gains, as evidenced in one of the datasets.
Given these gains and the proven computational complexity of the problem, the development of approximation algorithms or distributed solutions incorporating IC are of high relevance.

Concluding, the novel problem setting of optimal link
activation with interference cancellation we have introduced here provides new insights for system and protocol design in the wireless networking domain, as in this new context, strong interference is helpful rather than harmful. Thus, the topic calls for additional research on resource allocation schemes in scheduling and
routing that can take the advantage of the interference cancellation capability. Indeed, the LA setup studied herein assumes fixed transmit power for active links. This can lead to increased interference levels, since the SINRs can be oversatisfied. Incorporating power control to the LA problem with IC will bring another design dimension which can yield additional gains. 
Furthermore, it may enable IC for high SINR thresholds IC.

\end{document}